\documentclass[submitting]{nst}

\usepackage{subfigure,dcolumn}
\usepackage{epstopdf}
\usepackage{mhchem}
\usepackage{upgreek}
\usepackage{graphicx}
\usepackage{xcolor}

\begin{document}

\title{Performance of compact plastic scintillator strips with WLS-fiber and PMT/SiPM readout}
\thanks{Supported by NSFC (No. 11875282 and 11475205)}

\author{Min Li}
\affiliation{Institute of High Energy Physics, Beijing 100049, China}
\affiliation{University of Chinese Academy of Sciences, Beijing 100049, China}
\author{Zhimin Wang}
\email[Corresponding author,]{wangzhm@ihep.ac.cn.}
\affiliation{Institute of High Energy Physics, Beijing 100049, China}
\author{Caimei Liu}
\affiliation{Institute of High Energy Physics, Beijing 100049, China}
\affiliation{University of Chinese Academy of Sciences, Beijing 100049, China}
\author{Peizhi Lu}
\affiliation{Sun Yat-sen University, Guangzhou 510275, China}
\author{Guang Luo}
\affiliation{Sun Yat-sen University, Guangzhou 510275, China}
\author{Yuen-Keung Hor}
\affiliation{Sun Yat-sen University, Guangzhou 510275, China}
\author{Jinchang Liu}
\affiliation{Institute of High Energy Physics, Beijing 100049, China}
\author{Changgen Yang}
\affiliation{Institute of High Energy Physics, Beijing 100049, China}

\begin{abstract}
This work presents the design and performance study of compact strips of plastic scintillator with WLS-fiber readout in a dimension of $0.1\times0.02\times2\,m^3$, which evaluates as a candidate for cosmic-ray muon detector for JUNO-TAO. The strips coupling with 3-inch PMTs are measured and compared between the single-end and double-end readout options first, and the strip of double-end option coupling with SiPM is further measured and compared with the results of that with the PMTs. The performance of the strips determined by a detailed survey along their length with cosmic-ray muon after a detailed characterization of the used 3-inch PMTs and SiPMs.  The proposed compact strip of plastic scintillator with WLS-fiber coupling with SiPM provides a good choice for cosmic-ray muon veto detector for limited detector dimension in particular.
\end{abstract}


\maketitle

\section{Introduction}
\label{1:intro}

Muon flux reaching the surface of the Earth makes the most abundant cosmic-ray–induced radiation at sea level\cite{doi:10.1080/00295450.2018.1522885}. It has studied and utilized after the discovery by Anderson and Neddermeyer at Caltech in 1936\cite{AndersonCarlD1936CCOo}. To tag and veto cosmic muons with highly efficiency, veto systems are crucial for low background experiments such as searching for  neutrino\cite{2022103927,PROSPECT-ASHENFELTER2019287}, Dark Matter\cite{osti_1250509,PandaX-4T2019} and Double Beta decay\cite{EXO-200:2011xzf}. Generally, we can accomplish the discrimination of muon using two methods: the expected energy deposition with a simple energy threshold and a coincidence measurement. Requirements for such a muon tag and veto system are high efficiency for muon identification, immunity from the ambient gamma-ray background, size in a limited detector dimension, and low cost per mass unit\cite{GIANNINI2001316}. It is important for experiments with limited overburden or even deep underground experiments.

The Taishan Antineutrino Observatory (JUNO-TAO)\cite{TAO2020arXiv200508745J} is a satellite experiment of the JUNO experiment\cite{JUNOphysics}, a ton-level liquid scintillator (LS) detector placed at $\sim$30 meters from a reactor core of the Taishan Nuclear Power Plant in Guangdong, China. The main purposes of TAO are to provide a reference antineutrino spectrum for JUNO to remove model dependencies in the determination of the neutrino mass ordering, and to provide a benchmark measurement to test nuclear databases. A compact and high-efficiency muon detector is needed to suppress the muon-related background, where the location of the TAO detector near the reactor only has limited space, limited overburden, and a higher muon flux. Following the proposal of the JUNO-TAO detector\cite{TAO2020arXiv200508745J}, a multi-layer detector of the plastic scintillator is proposed as a muon tag and veto detector to cover around 4\,m$\times$4\,m on the top of TAO detector.

Many kinds of detectors use for detecting muons. High-efficiency organic plastic scintillation (PS) detectors are widely applied as a proven technology for their excellent optical transmission properties, simple production, low cost, stability, fast response time, sensitivity to all kinds of radiation, and an excellent capacity to handle high-radiation background environments. 
In lots of high-energy physics projects, plastic scintillator strips served as an anti-coincidence detector to provide a trigger signal and are applied as sensitive elements for tracking detector, such as OPERA\cite{Adam2007TheOE-PS}, MINOS\,\cite{MINOS-PS}, the K2K SciBar detector\cite{NITTA2004147}, Minerva\cite{MINERvA-ps}, TAE\cite{TEA2019}, AugerPrime\cite{ZONG201882}, \textmu Cosmics\cite{T2Cosmics2019}, YBJ-HA\cite{LHAASO-KM2A2018}, LHAASO \cite{LHAASO-ED2021,AHARONIAN2021165193}, muon tomography\cite{X-ray6551311,PARK2021,2018Position} and many other applications\cite{Position-Sensitive4033267,HOU2019102727}. 

In general, the light yield (LY)\cite{SENCHYSHYN2007911,PLADALMAU2001482} of a scintillator, and the detection efficiency, are the key criterion to describe the quality of the detection set-up\cite{ZONG201882}.
Excellent uniformity and relatively high light collection are required to achieve high efficiency for muons while maintaining good discrimination from gammas. Wavelength shifting (WLS) fibers coupled with a photomultiplier (PMT) or multi-pixel silicon-based avalanche photo-diodes operated in Geiger mode (SiPM) are commonly used to avoid bulky light guides and read out the light from scintillators\cite{BUGG201491,seo2022feasibility,Mineev2011}. The PS can be much more mechanically robust and offer great flexibility in detector size and shape, better tolerance to magnetic fields, higher photon detection efficiency, compactness, and low cost with SiPM in particular\cite{doi:10.1080/00295450.2018.1522885,Garutti_2011,Papa_2014}. 
Normally, the WLS fibers are placed into grooves or holes along the strip, and the detection efficiency can be significantly increased by improving the optical contact between the scintillator and the fiber by adding an optical filler into the groove/hole with an optical glue having high transparency and a refractive index close to the refractive index of the strip base material (usual polystyrene), leading to a light yield increase of up to 50\% in comparison to the strips without a filler\cite{2016chep.confE.789D,ARTIKOV201987}.

In this study, we proposed a basic design on compact strips of PS with WLS-fiber in the dimension of $0.1\times0.02\times2m^3$, aiming for a compact muon tagging detector with good efficiency and identification of gammas. Two strips with 1\,mm WLS-fiber as prototype are fabricated and tested. The comparisons among the readout options of single-end or double-end, and sensor options of PMT or SiPM are done in detail by cosmic-ray muon survey. Sec.\,\ref{2:setup} will introduce the design of the strips and the testing system. Sec.\,\ref{3:results} will show the results in detail. Finally, a summary will be provided in Sec.\,\ref{4:summary}.

\section{System Setup}
\label{2:setup}

Following the studies and strategies discussed in Sec.\,\ref{1:intro}, two kinds of strips of PS with WLS-fiber as a prototype are designed and produced for R\&D. In this section, the strip design and the testing system will be introduced. The used photon sensors of 3-inch PMT or SiPM also will be characterized.

\subsection{PS strip with WLS-fiber}
\label{2:setup:1:ps}

\begin{figure*}[!htb]
\begin{center}
\subfigure[Design of PS with WLS-fiber]{
\includegraphics[width=3.2in,height=2.4in]{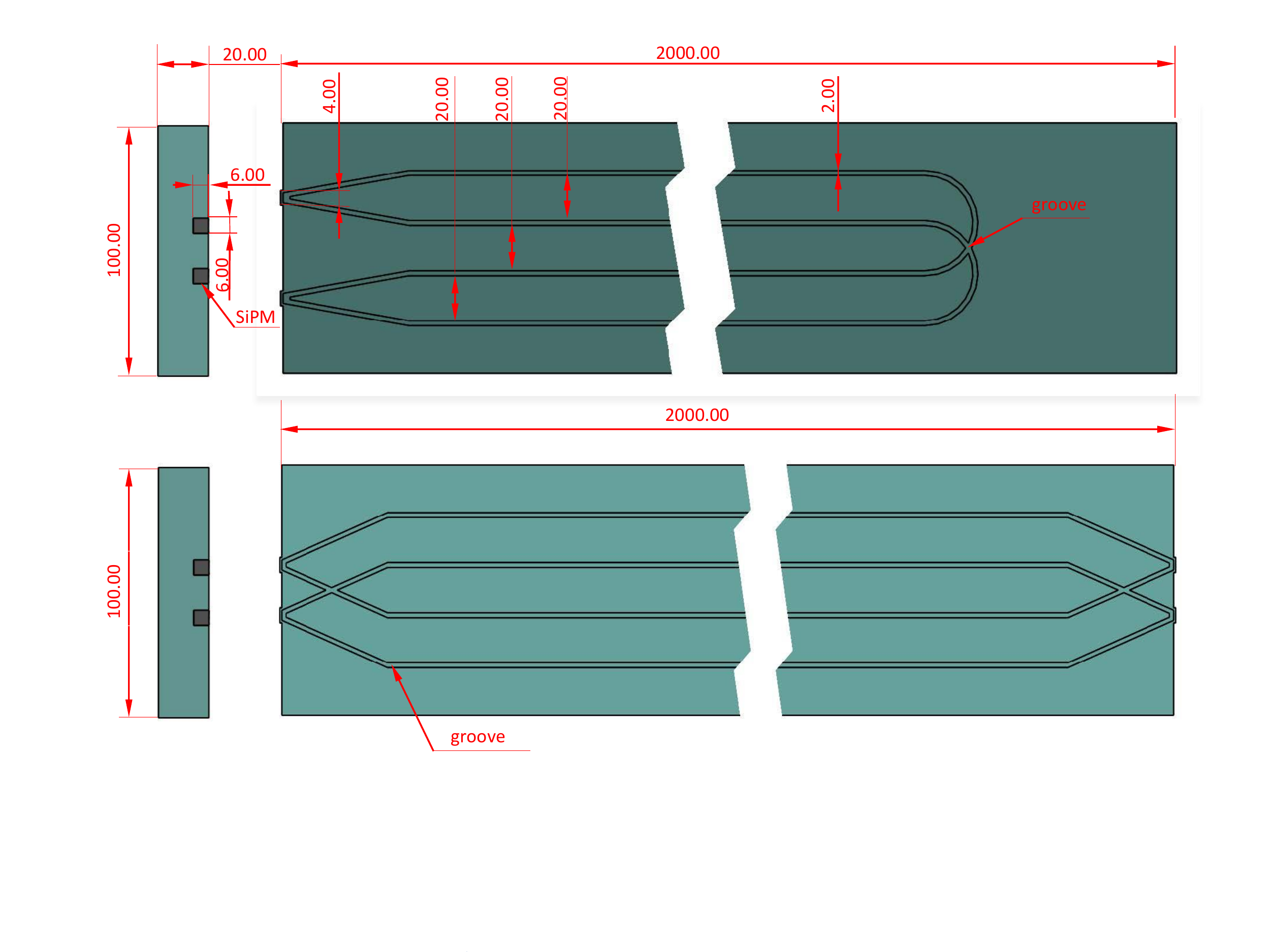}
\label{fig:ps-module1}
}
\subfigure[Gathered fibers and optical window]{
\includegraphics[width=3.2in,height=2.4in]{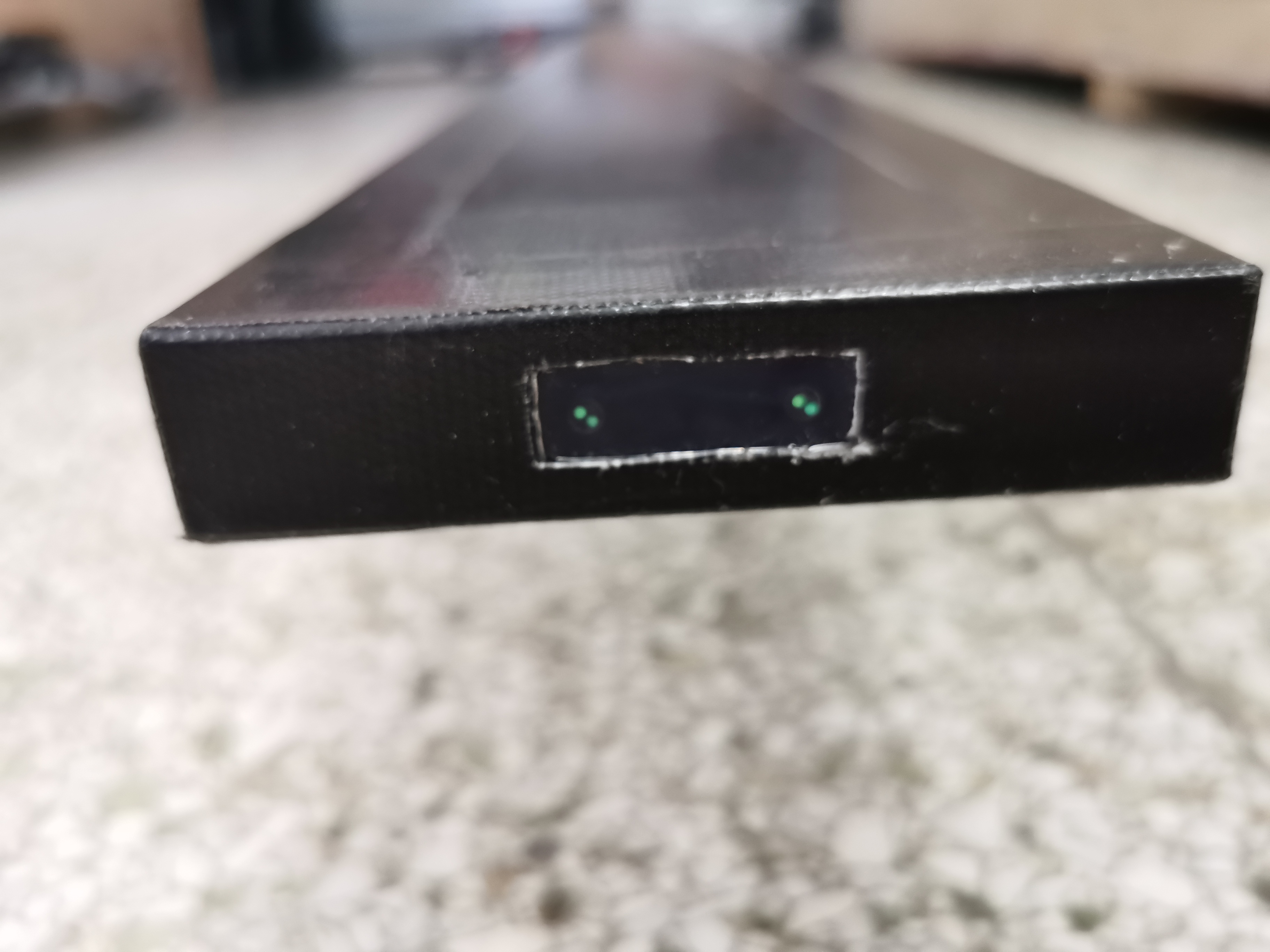}
\label{fig:ps-module2}
}
\caption{Comparison of light yield and muon efficiency of PS strips with SiPMs and PMTs. green: double-end (SiPM), orange: double-end (SPMT), blue: single-end (SPMT)}
\label{fig:ps-module} 
\end{center}
\end{figure*}

The design of the compact PS strips with WLS-fiber is proposed with two readout options of single-end or double-end shown in Fig.\,\ref{fig:ps-module1}, where the key features are the filled gaps between the scintillator and the fiber, and the flat surface of the fibers to the PS at its end (Fig.\,\ref{fig:ps-module2}). The dimension of the PS strip is $0.1\times0.02\times2m^3$ (Width$\times$Thickness$\times$Length), with four 1\,mm WLS-fibers along its width direction (around 2\,cm spacing between neighbor fibers) which are inserted and glued by a filler into the grooves on the surface of the PS. The four fibers of the double-end option will be gathered into two groups at each of the ends (four groups in total) and coupled to the photon sensors. There will be only two groups at the output end of the single-end option (no fiber cut at another end) aims to reduce the sensor and electronics channels. 

The prototypes of the designed PS strips are finalized and fabricated by \textit{Beijing Hoton Nuclear Technology Co., Ltd.} \cite{hoton-web}. The two strips are made with extruded plastic scintillator type of SP101 polymerized with liquid polystyrene added by P-triphenyl and POPOP. The groove is designed to $6\times6\,mm^2$ in an optical window of $1\times4\,cm^2$ as shown in Fig.\,\ref{fig:ps-module2}, where the WLS-fiber BCF92\cite{Saint-Gobain-web} with diameter 1\,mm is used and its end surface is flat. The PS strip is covered with 0.08\,mm Al film firstly, then another 0.8\,mm PVC layer, and finally packaged by a black adhesive tape layer. The scintillation photons will be collected through the WLS-fiber and read out by photon sensors such as SiPM or PMT, which will be measured and compared in detail and discussed in the following sections.

\subsection{Electronics and DAQ}
\label{2:setup:2:elec}

\begin{figure*}[!htb]
\begin{center}
\subfigure[Double-end setup with SPMTs]{
\includegraphics[width=3.2in,height=2.4in]{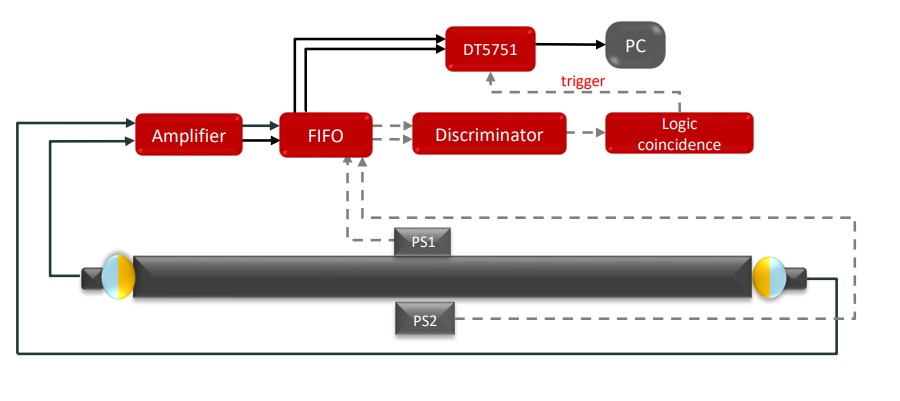}
\label{fig:spmtsetup:doubleend}
}
\subfigure[Single-end SPMTs and double-end SiPMs]{
\includegraphics[width=3.2in,height=2.4in]{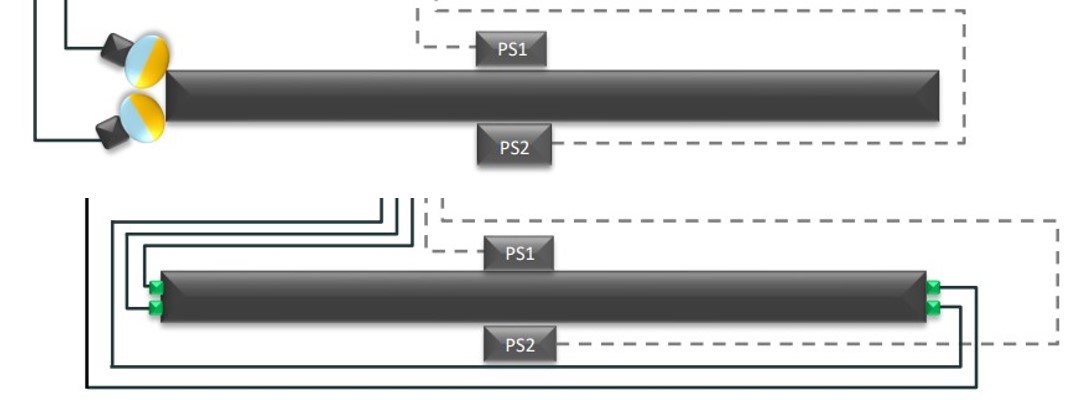}
 \label{fig:spmtsetup:short}
}
\caption{Schema of system setup. The mini-modules (PS1 and PS2) will survey nine locations of the PS strip with 0.2\,m step between -0.8\,m (minimum -1\,m, left end, side A) and 0.8\, m (maximum 1\, m, right end, side B)}
 \label{fig:setup:schema}
\end{center}
\end{figure*}

\begin{figure*}[!htb]
\begin{center}
\subfigure[The style of coupling]{
\includegraphics[width=3.2in,height=2.4in]{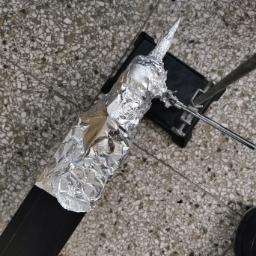}
\label{fig:coupling}
}
\subfigure[trigger]{
\includegraphics[width=3.2in,height=2.4in]{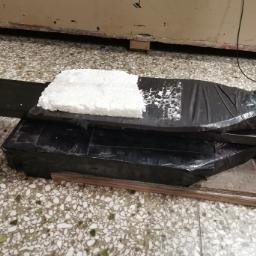}
\label{fig:trigger}
}
\caption{Experimental setup}
\label{fig:setup}
\end{center}
\end{figure*}

A general schema of the testing system can be found in Fig.\,\ref{fig:setup:schema}. Two small PS modules (mini-module, PS1 and PS2) locate above and below the PS strip, respectively (Fig.\,\ref{fig:trigger}), which will use as muon tagging. The signals of the 2-inch PMTs of the PS mini-modules send into a FIFO module (OCTAL LINEAR model 748) and then discriminated by a low threshold discriminator (CERN N845), the coincidence of which (CAEN logic module N455) used to tag a muon as a trigger of the data taking system. The mini-modules will survey along the length of the PS strip for uniformity checking, which will use for all the following tests single-end, double-end, PMT and SiPM. 
The two 3-inch PMTs (SPMT) or four SiPMs use to collect the photons, where the SiPMs (SPMTs) are coupled to the PS strip through the air directly (Fig.\,\ref{fig:coupling}). 
Please note that the two SPMTs are used for both single-end and double-end options: one SPMT for one end (two fiber groups) of double-end, and one SPMT for one fiber group of single-end (two SPMTs used in total, Fig.\,\ref{fig:spmtsetup:short}). Each of the SiPMs (four in total) covers one fiber group of the double-end option (Fig.\,\ref{fig:spmtsetup:short}).
All the signals of the sensors are sent into the FIFO first and then recorded in waveforms by an FADC (CAEN DT5751, 1\,GS/s, 1\,V p-p) triggered by the mini-modules. 
Each SiPM (SPMT) will have its own individual HV power supply, and an additional amplifier is applied to SiPM4 to improve its signal-to-noise ratio discussed later.

\subsection{Mini-modules for muon trigger}
\label{2:setup:3:ps:mini}

\begin{figure*}[!htb]
\begin{center}
\subfigure[Amplitude]{
\includegraphics[width=3.3 in,height=2.4in]{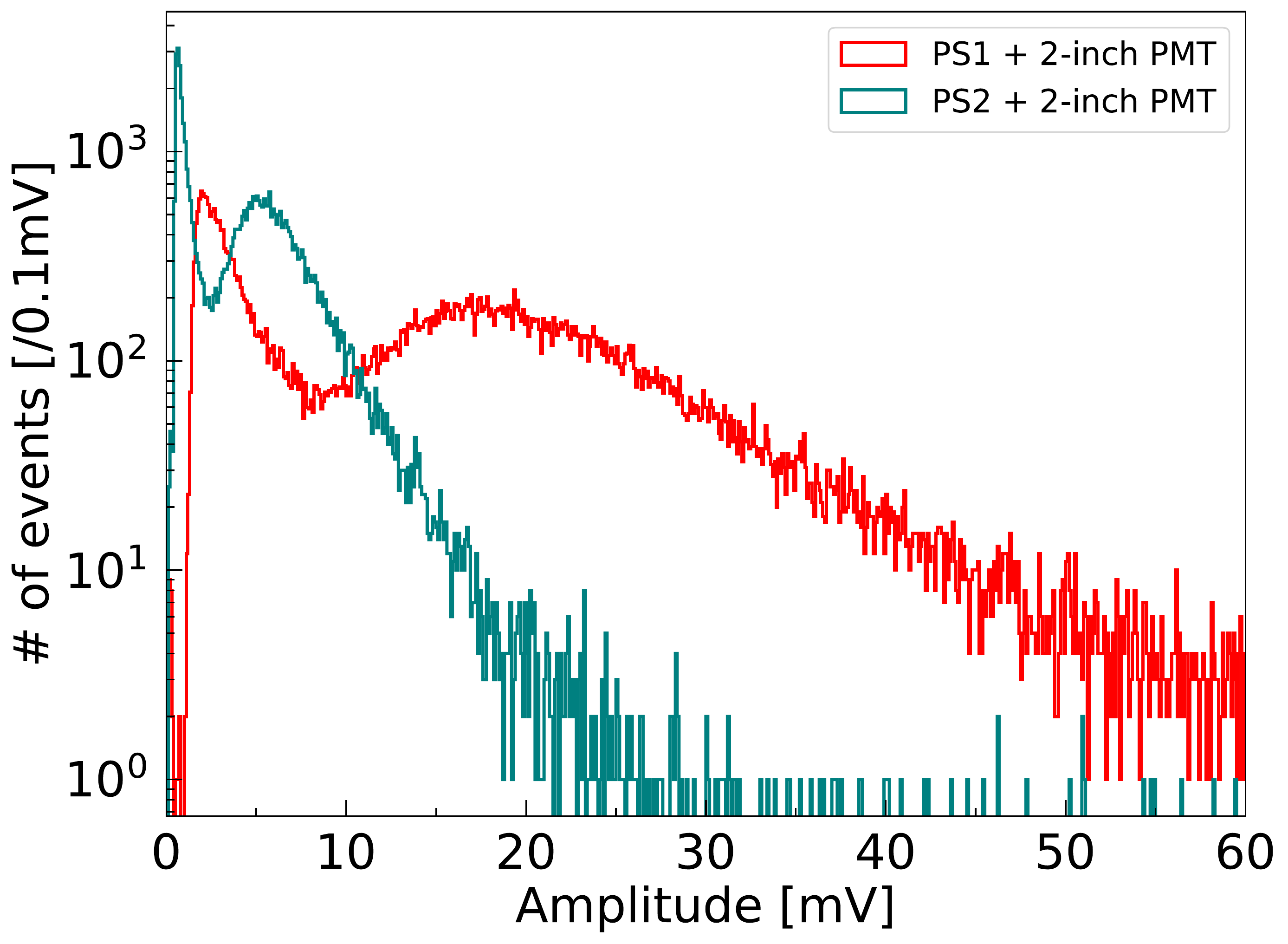}
\label{fig:ps:minimodule:a}
}
\subfigure[Charge]{
\includegraphics[width=3.2in,height=2.4in]{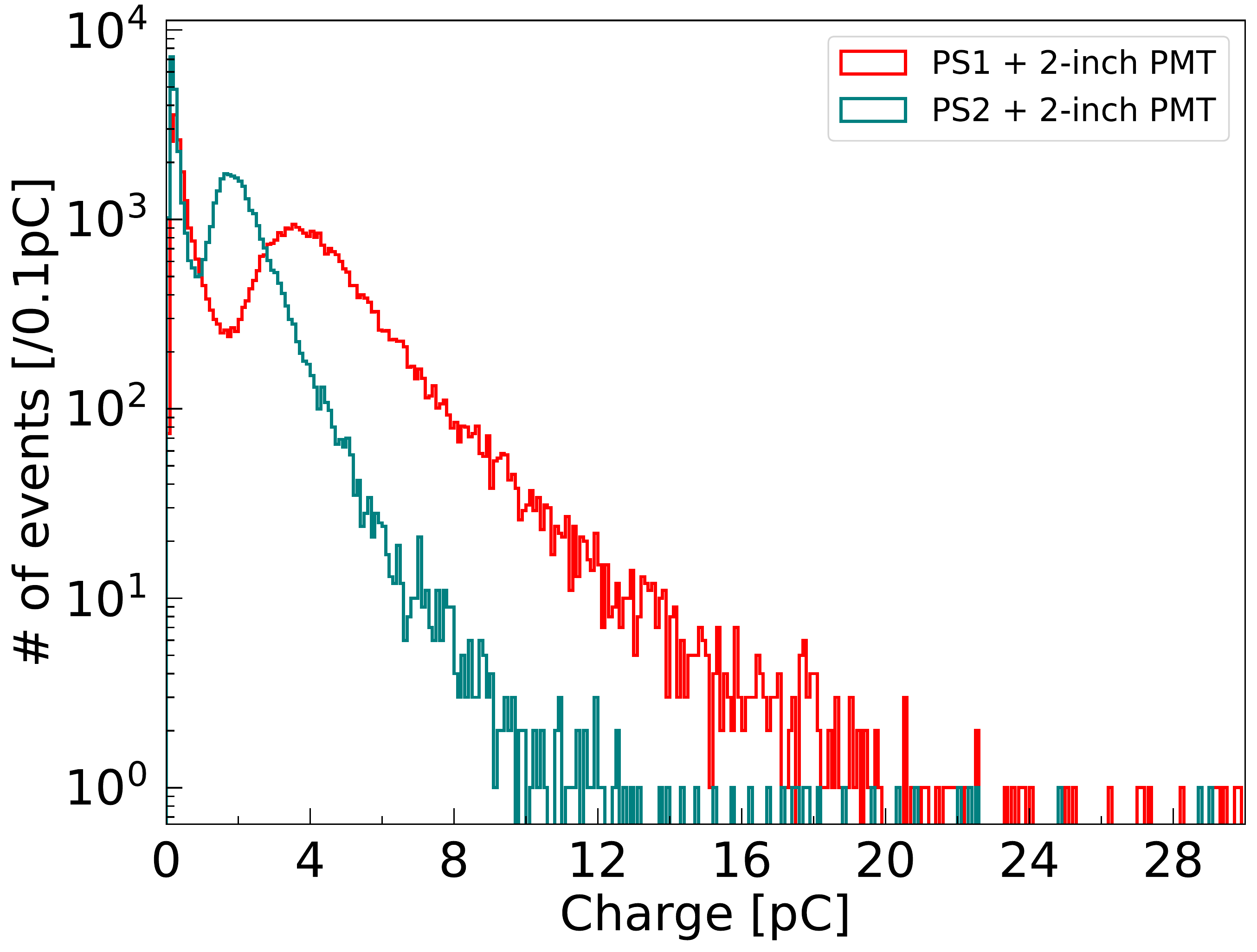}
\label{fig:ps:minimodule:b}
}
\caption{Amplitude and charge spectra of the mini-modules}
\label{fig:ps:minimodule} 
\end{center}
\end{figure*}

The two mini-modules (PS1, $25\times4\times30\,cm^3$ and PS2, $15\times1\times22\,cm^3$), equipped with 2-inch PMT, are used to tag muons and calibrate the performance of the designed PS strips. The measured spectra of the mini-modules in amplitude and charge are shown in Fig.\,\ref{fig:ps:minimodule}, where the 2-inch PMTs set to -2200\,V (PS1) and -1800\,V (PS2) for a similar operating gain, respectively. The threshold of the mini-modules in the amplitude sets to around 10\,mV (PS1) and 2\,mV (PS2) (Fig.\,\ref{fig:ps:minimodule:a} to select muons in the following measurements. The value difference of the threshold is mainly from their light yield related to their thickness, which introduces different valley locations on their charge spectra shown in Fig.\,\ref{fig:ps:minimodule:b}. The coincidence rate of the two mini-modules is around 6-7\,Hz.

\subsection{SPMT and SiPM}
\label{2:setup:3:sensor}

The PS strips will be coupled with SiPMs and SPMT to collect the photons around room temperature (25$^{\circ}$C). Four pieces of SiPMs will use in total here, including three of HPK S14160 in the dimension of $3\times3\,mm^3$\cite{HPK-SiPM-web} (SiPM1, SiPM2, SiPM3), which claims a higher PDE (50\% at $\lambda_p$) and lower operation voltage, and another from S12572 (SiPM4). The 3-inch PMTs (SPMT) from HZC\cite{JUNO3inchPMT} is used as a reference to compare with the SiPMs, for single-end and double-end options. The characterization of SiPM is done firstly with a similar procedure as in \cite{SiPMtest-KLANNER201936} but mainly focuses on its gain versus over-voltage (OV) by LED, cross-talk (CT), and dark count rate (DCR). 

\begin{figure*}[!htb]
\begin{center}
\subfigure[Charge spectra of SiPMs]{
\includegraphics[width=3.2in,height=2.4in]{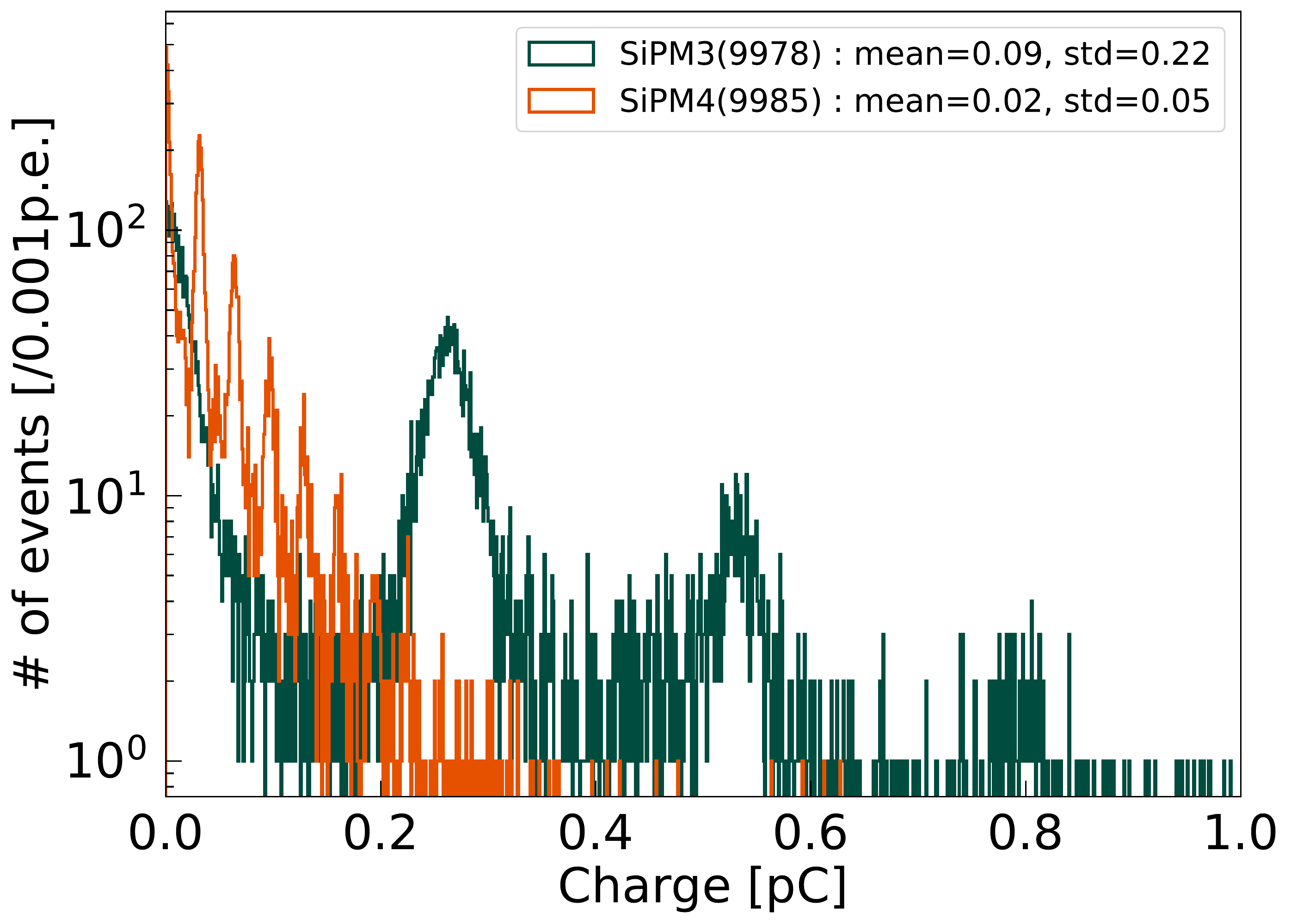}
\label{fig:sipm:charge}
}
\subfigure[Gain vs. OV]{
\includegraphics[width=3.2in,height=2.4in]{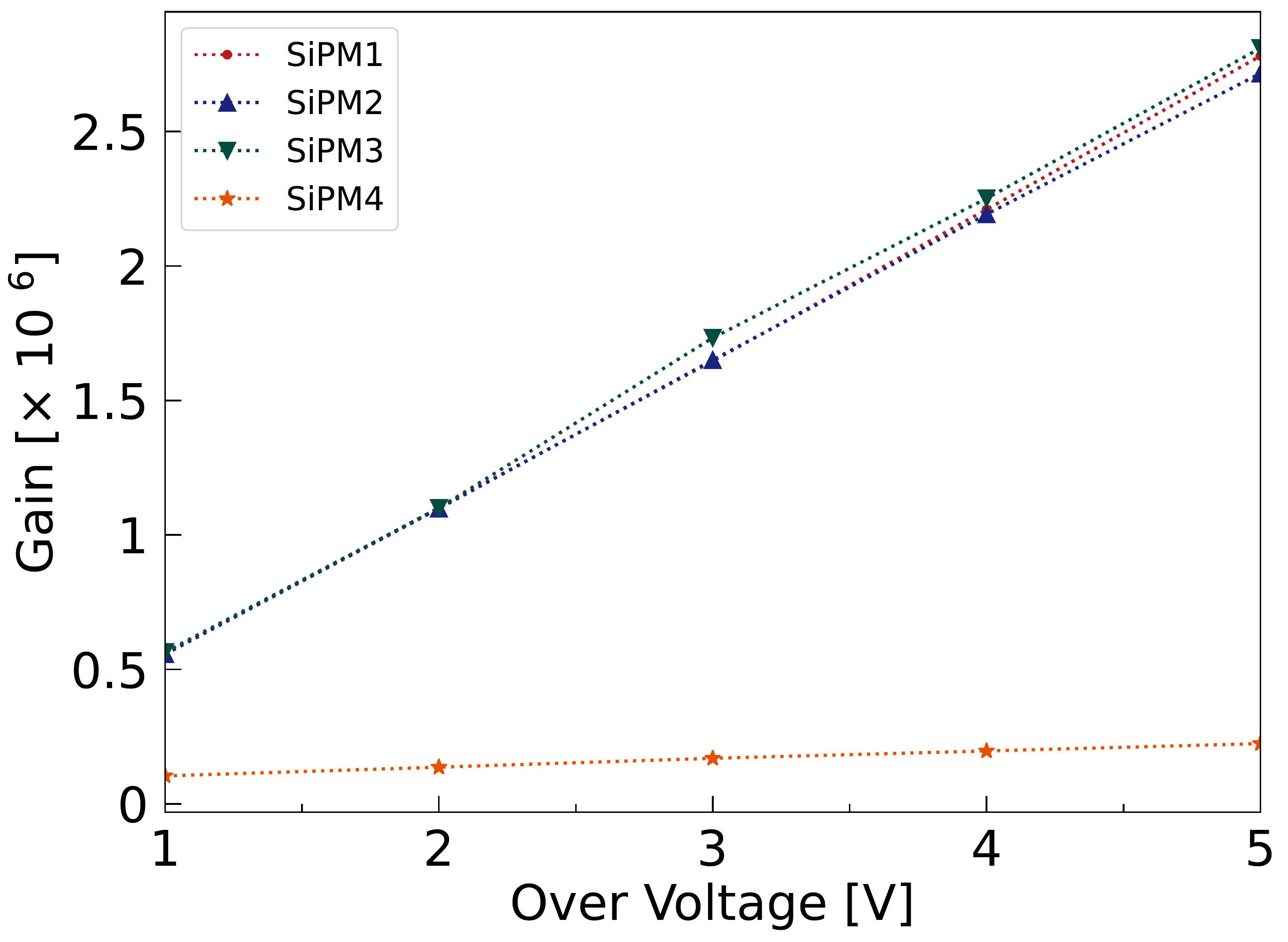}
\label{fig:gainvsov}
}
\newline
\subfigure[DCR vs. amplitude]{
\includegraphics[width=3.2in,height=2.4in]{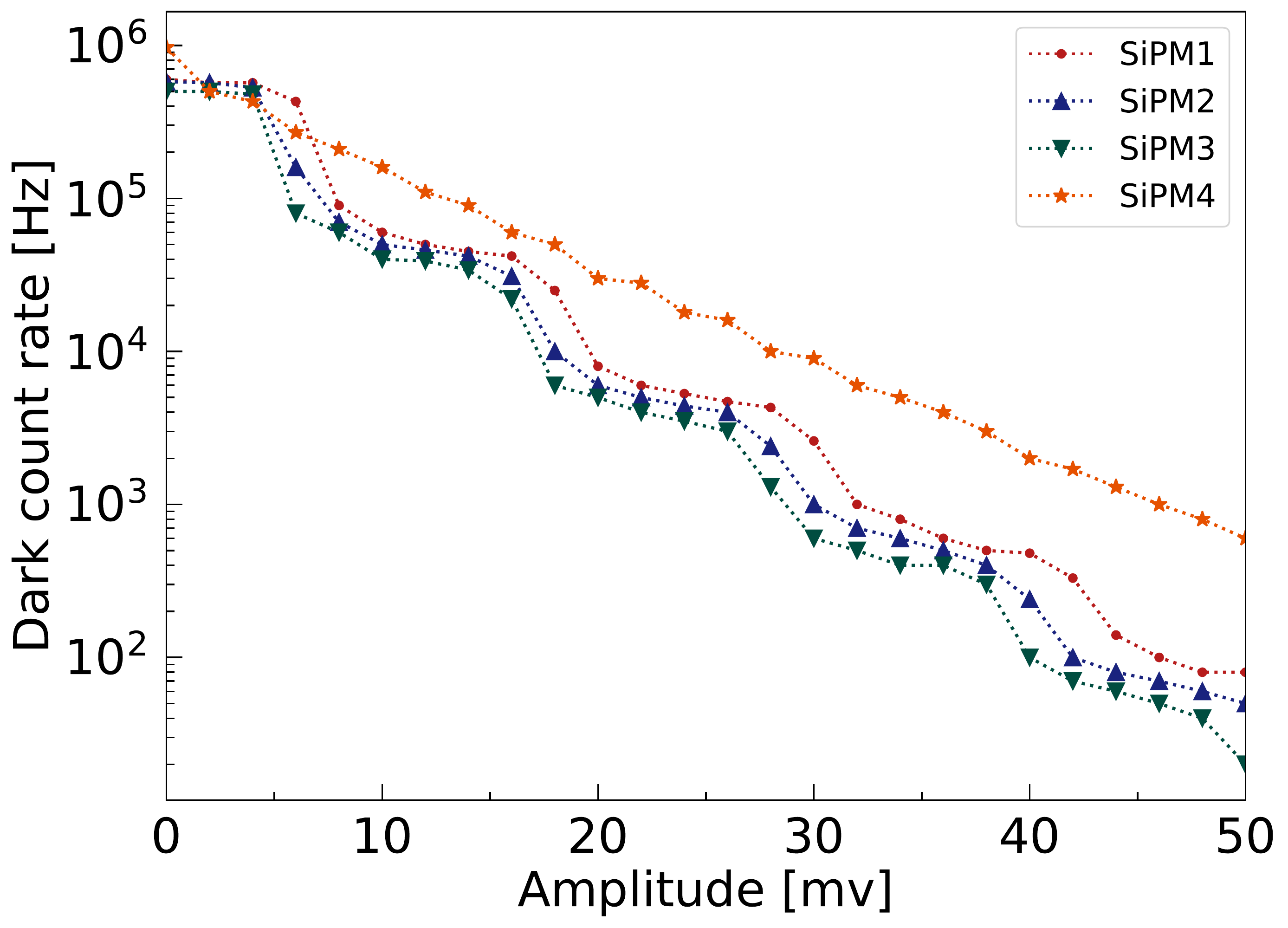}
\label{fig:dcrvsamplitude}
}
\subfigure[CT vs. OV]{
\includegraphics[width=3.2in,height=2.4in]{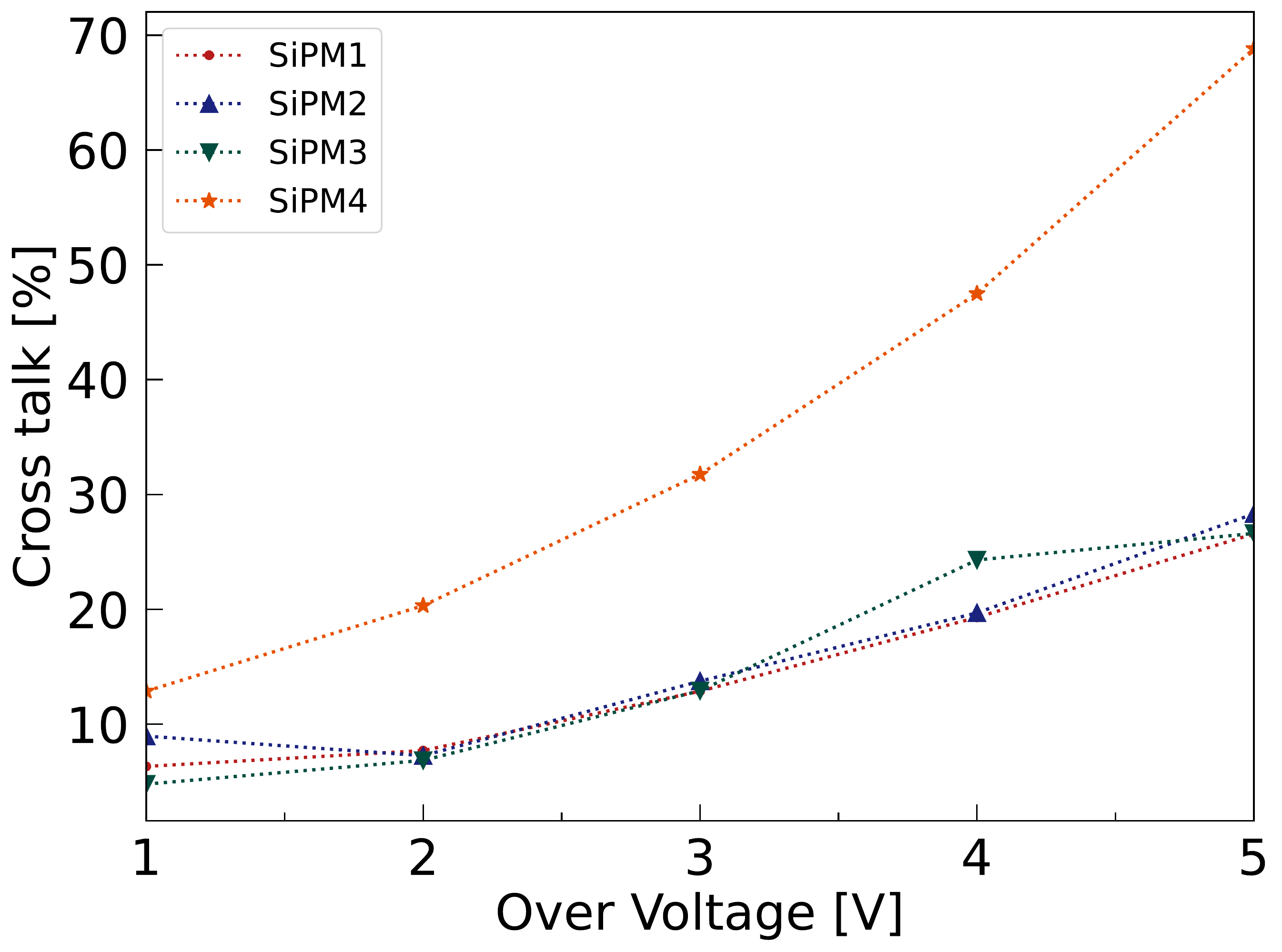}
\label{fig:ct}
}
\caption{Measured spectra of the SiPMs}
\label{fig:sipmcal} 
\end{center}
\end{figure*}

The measurements of SiPMs are shown in Fig.\ref{fig:sipmcal}, where the typical charge spectra measured with SiPM3 and SiPM4 are shown in Fig.\,\ref{fig:sipm:charge}, and the working gain of the SiPMs can calculate according to the peaks. The relationship of the gain versus OV is plotted in Fig.\,\ref{fig:gainvsov}, where the breakdown voltage (V$_{bd}$) is estimated by a fitting to the curve of gain versus applied voltage, where it is around 41\,V for SiPM1-3 and 71\,V for SiPM4. The OV of 3\,V sets to all the SiPMs. The SiPM uses another fast $\times10$ amplifier because of its low gain. The typical plot of DCR versus amplitude threshold of SiPM is shown in Fig.\,\ref{fig:dcrvsamplitude}, where the DCR is decreasing in steps by increasing the threshold as expected. It is around 500\,kHz at a half p.e. equivalent threshold 55\,kHz/mm $^2$). The measured cross-talk (CT) ratio is around 12\% for SiPM1-3 and 46\% for SiPM4. The very high cross-talk ratio of SiPM shown in Fig.\,\ref{fig:ct} will affect its charge measurement. A decreasing factor of 1/8 was applied to estimate the DCR for every other p.e. threshold increasing.
The DCR and CT will also affect the threshold setting, muon efficiency, and random coincidence expectation of the PS strip.

\begin{figure*}[!ht]
\subfigure[SPE charge spectrum]{
\includegraphics[width=3.2in,height=2.4in]{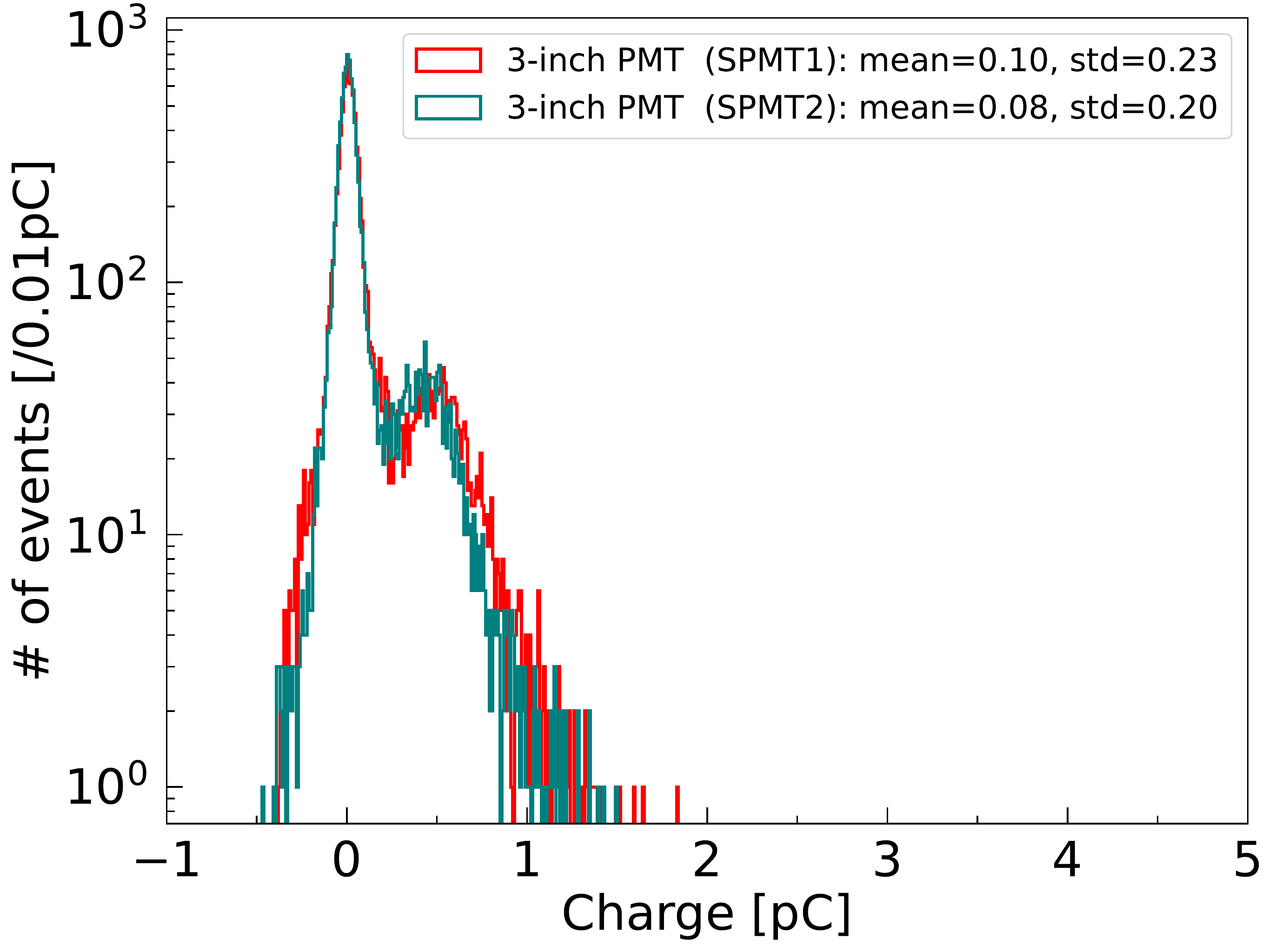}
\label{fig:spmtchargecal}
}
\subfigure[DCR versus amplitude threshold]{
\includegraphics[width=3.2in,height=2.4in]{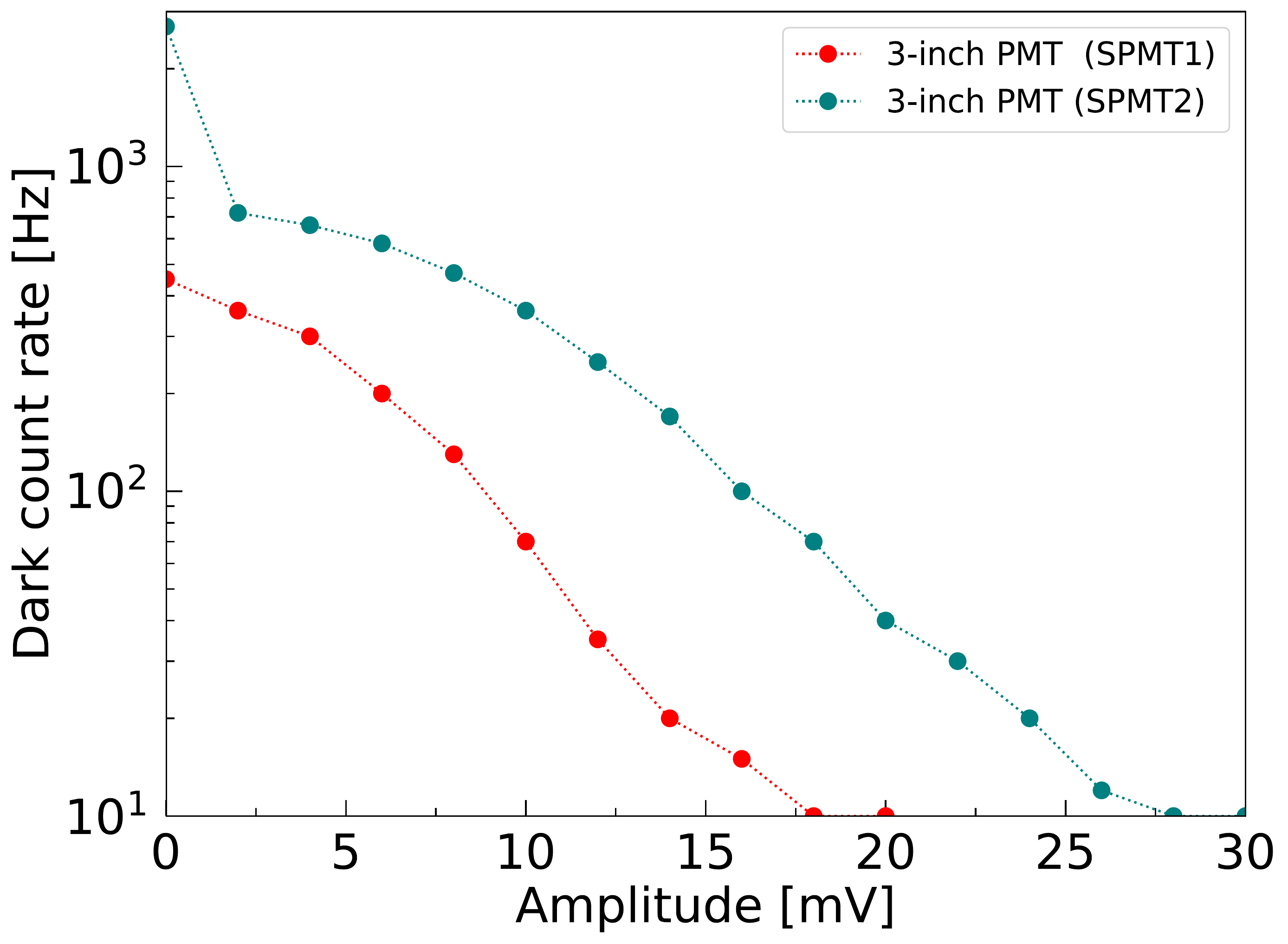}
\label{fig:spmtdcrvcamplitudecal}
}
\caption{Measured SPE charge spectrum and the DCR versus amplitude of the SPMTs}
\label{fig:spmt}  
\end{figure*}

The gain of the two used SPMTs is calibrated and tuned to $3\times$10$^6$ with positive HV 1150\,V and 1180\,V respectively. The QE of SPMTs is around 23\%, and DCR is around 400-700\,Hz at a single p.e.~threshold ($\sim$2.5\,mV/p.e.), which is much smaller than the SiPMs. The measured charge of single photo-electron (SPE) and the DCR versus threshold are shown in Fig.\ref{fig:spmt}. 

\section{Results and Discussion}
\label{3:results}

\begin{figure*}[!htb]
\subfigure[PMT hit-time of PS]{
\includegraphics[width=3.2in,height=2.4in]{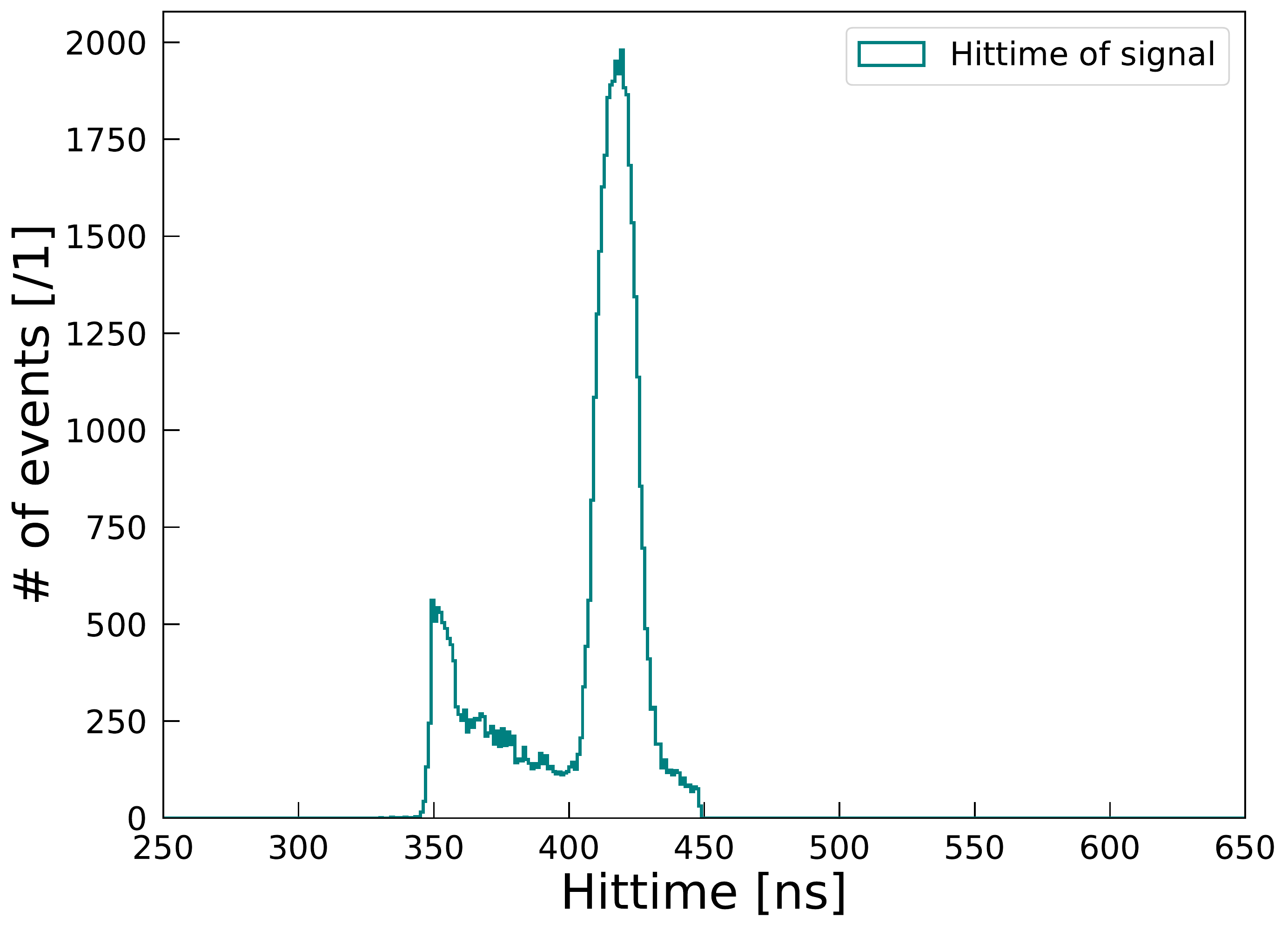}
\label{fig:hittimemuonsignal}
}
\subfigure[PMT light yield of PS]{
\includegraphics[width=3.2in,height=2.4in]{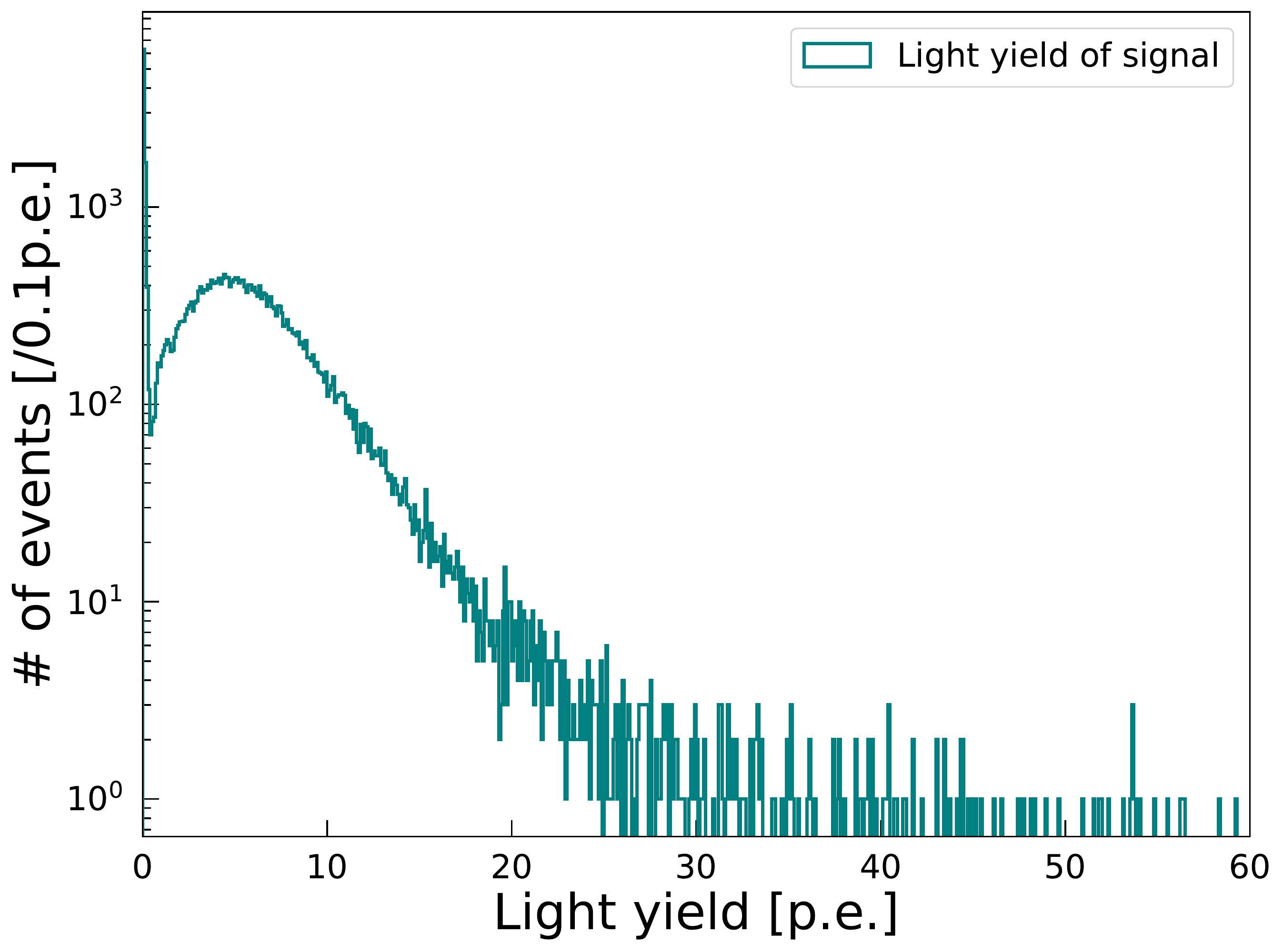}
\label{fig:spmtchargespectrumtrigger}
}
\caption{PMT charge and hit-time distribution of the PS strip with the coincidence trigger of the mini-modules.}
\label{fig:selectmuon}  
\end{figure*}

With the introduced system in Sec.\,\ref{2:setup}, the measurements and comparisons will be implemented for coupling with SiPM and PMT, and single-end and double-end options of the PS strips, respectively. The muon efficiency of the PS strips will be measured and calculated at nine different locations along its longitudinal direction. 
According to the measured charge and hit-time spectra of the PS strip triggered by the coincidence of the mini-modules as shown in Fig.\,\ref{fig:selectmuon}, further cuts on hit-time and charge are used to remove the random noise coincidence and select more pure muon samples hitting the PS strip, for example, hit-time in the range [390,420]\,ns for SPMTs ([320,360]\,ns for SiPMs) and charge higher than 0.4\,p.e., which will be used for the following analysis too. According to the measured charge spectrum, a systematic bias of the calculated muon efficiency is estimated to be less than 0.5\% from the applied charge threshold of 0.4\,p.e.

\subsection{PS strip with SPMT}
\label{3:results:1:PS:pmt}

\begin{figure*}[!htb]
\begin{center}
\subfigure[charge spectra of sing-end]{
\includegraphics[width=3.2in,height=2.4in]{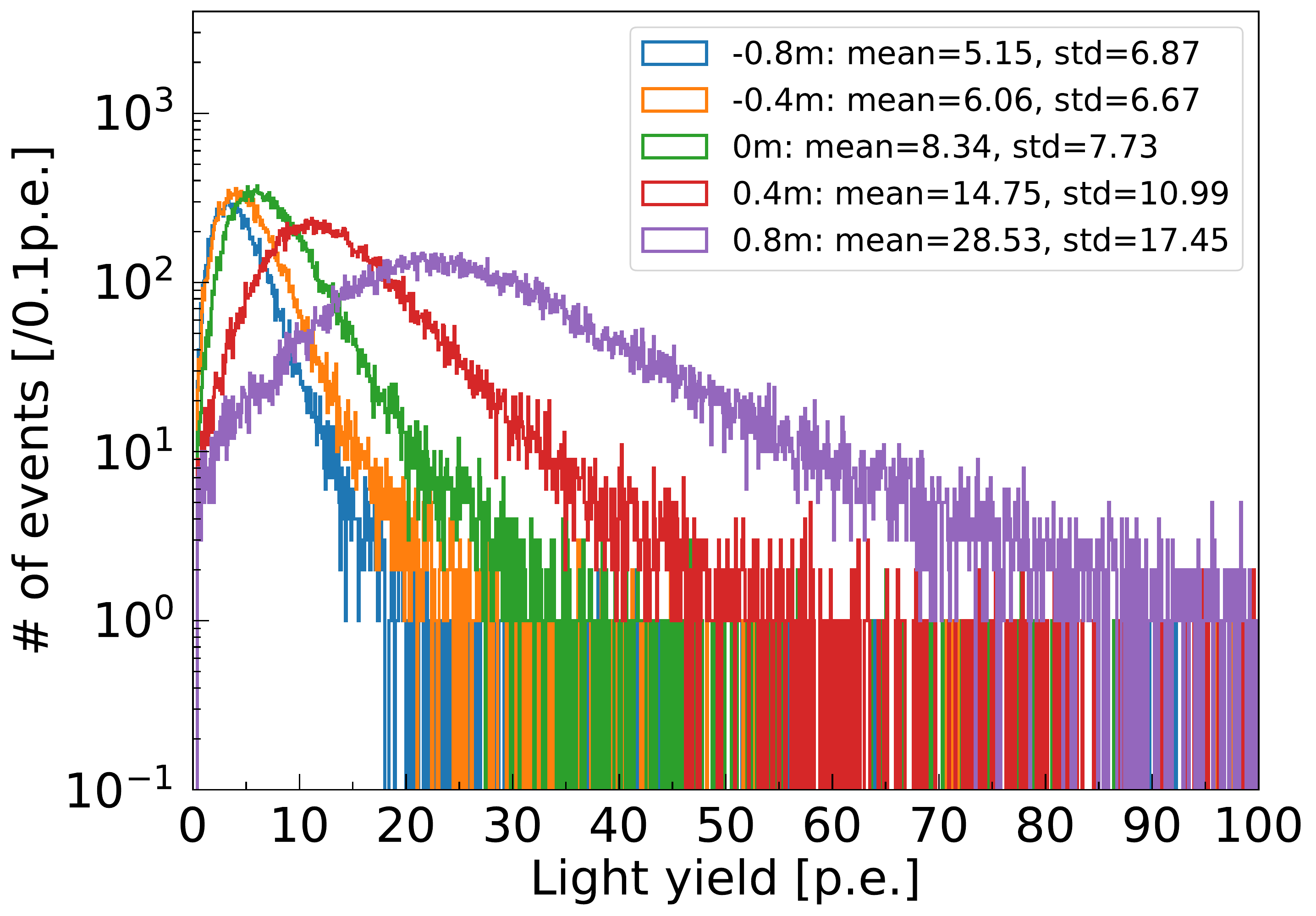}
\label{fig:spmtpecpectrum_1end}
}
\subfigure[charge spectra of double-end]{
\includegraphics[width=3.2in,height=2.4in]{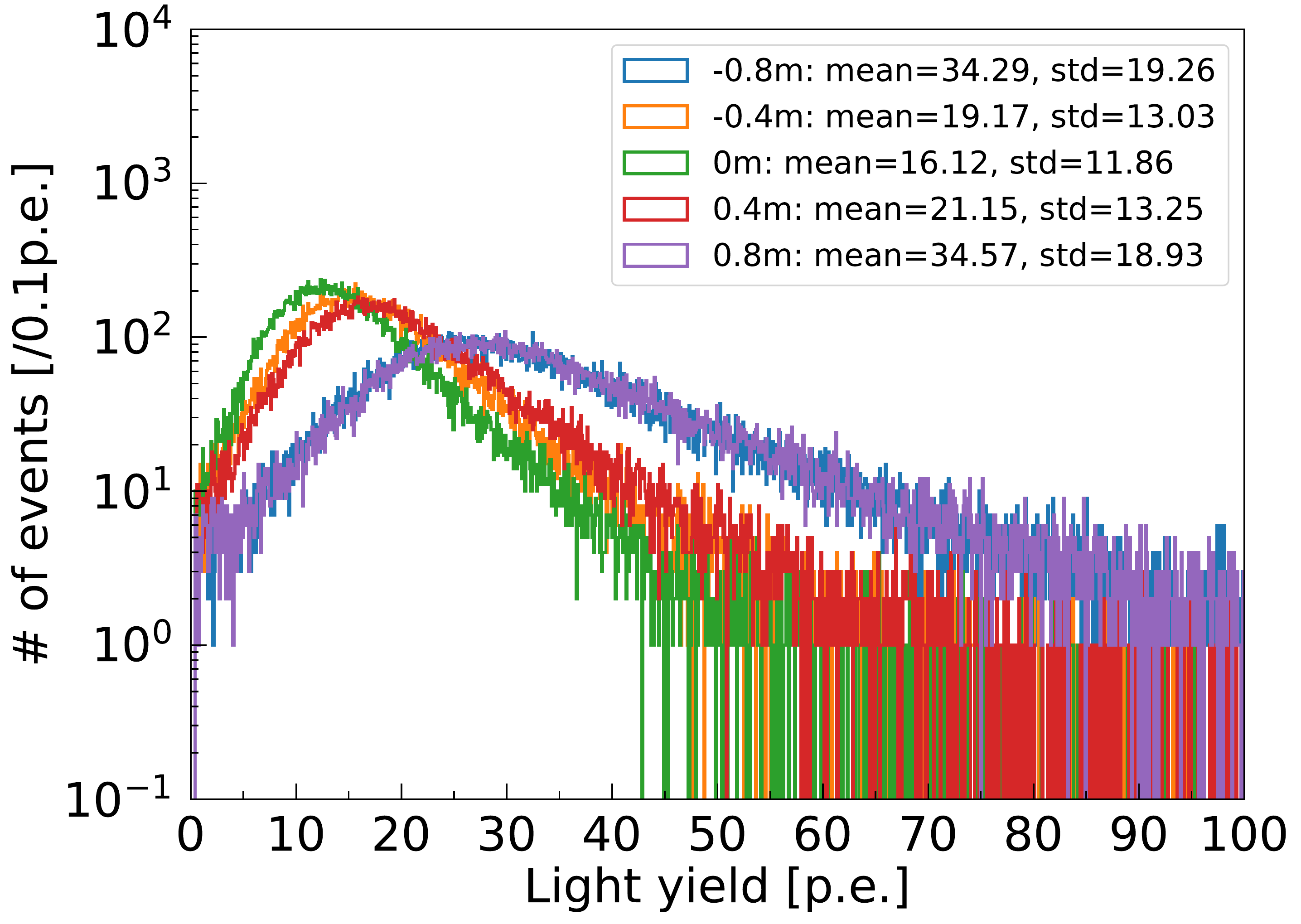}
\label{fig:spmtpecpectrum_2end}
}
\newline
\subfigure[LY vs. location of single-end]{
\includegraphics[width=3.2in,height=2.4in]{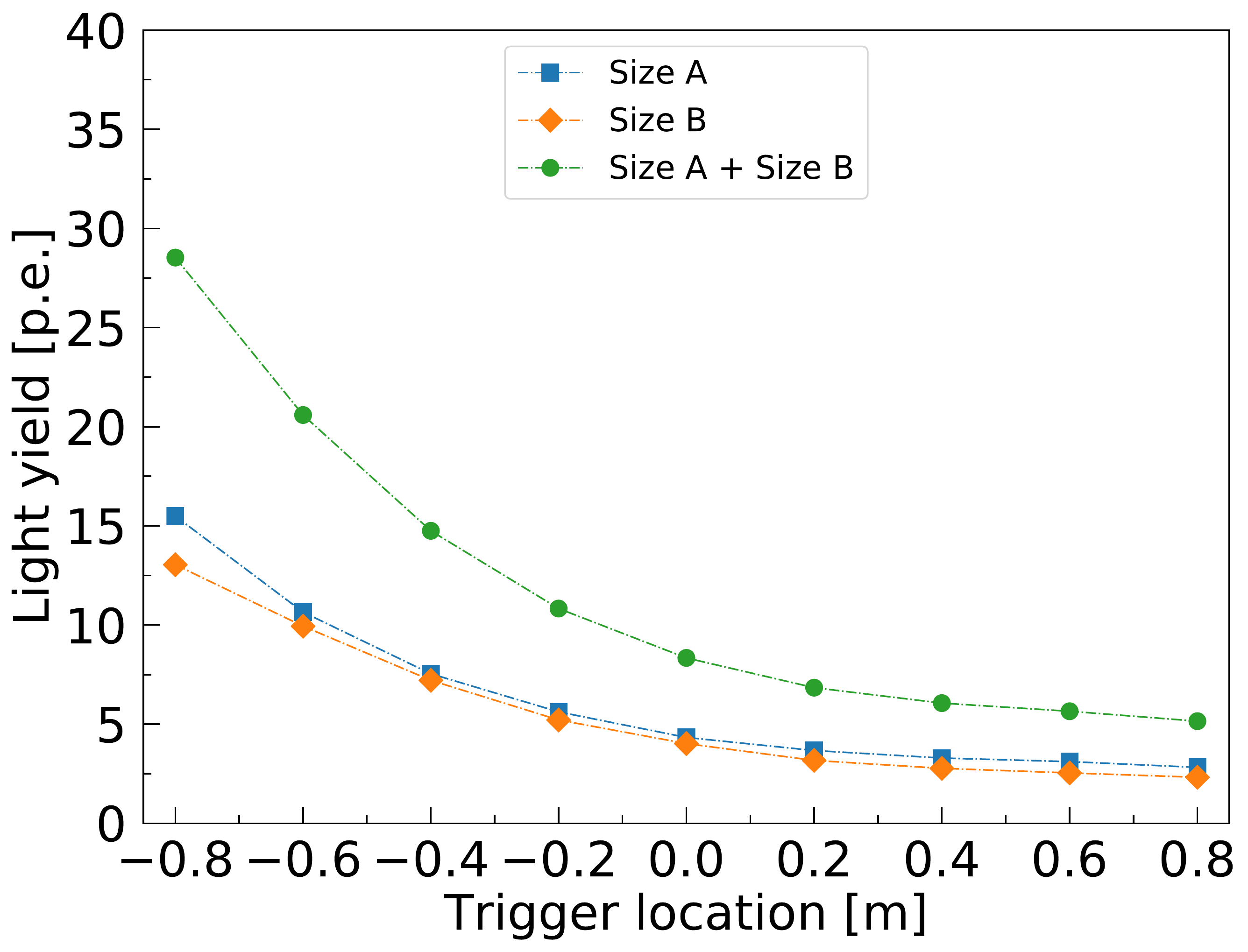}
\label{fig:spmtlightyield1}
}
\subfigure[LY vs. location of double-end]{
\includegraphics[width=3.2in,height=2.4in]{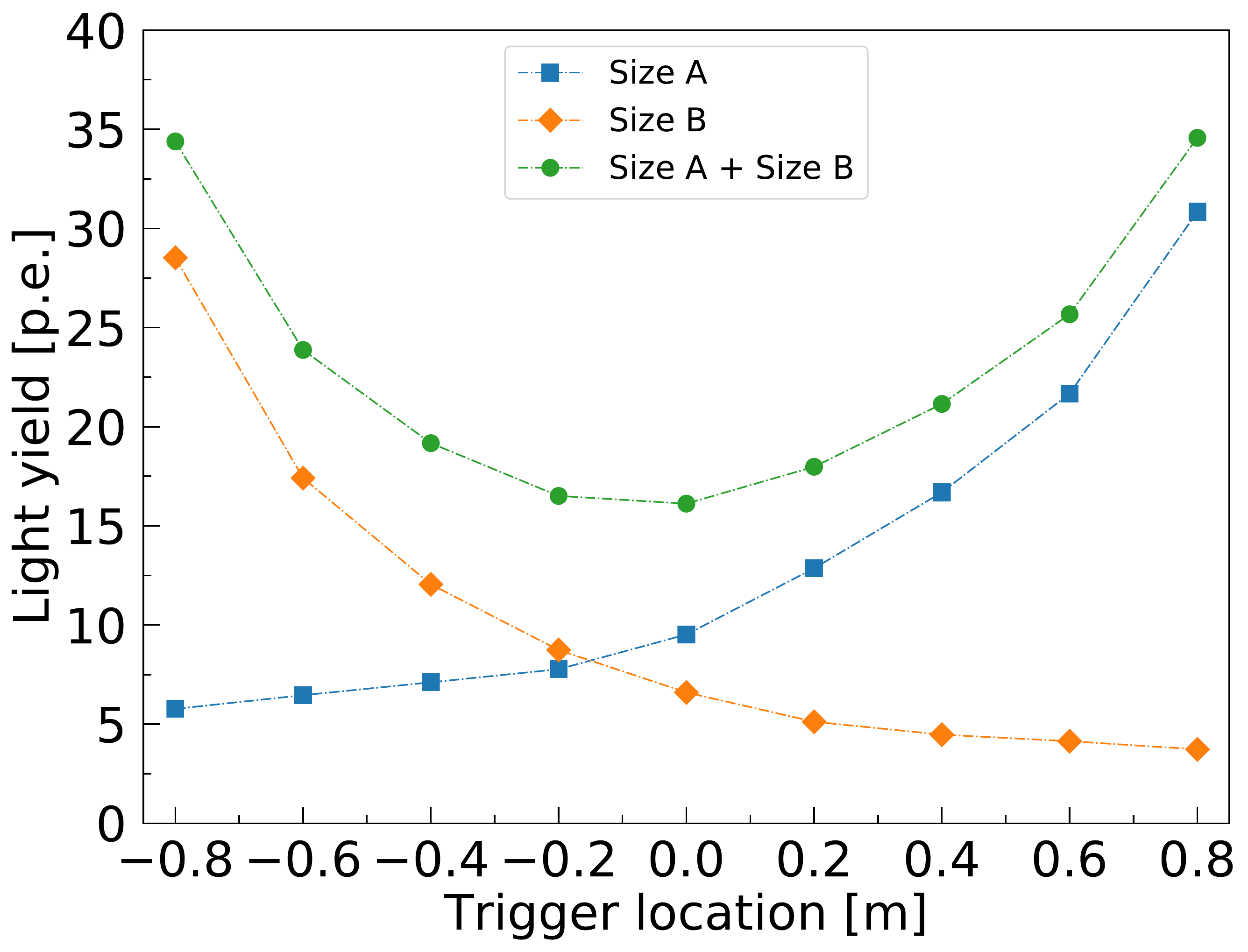}
\label{fig:spmtlightyield2}
}
\newline
\subfigure[Muon eff. vs. thrd. of single-end]{
\includegraphics[width=3.2in,height=2.4in]{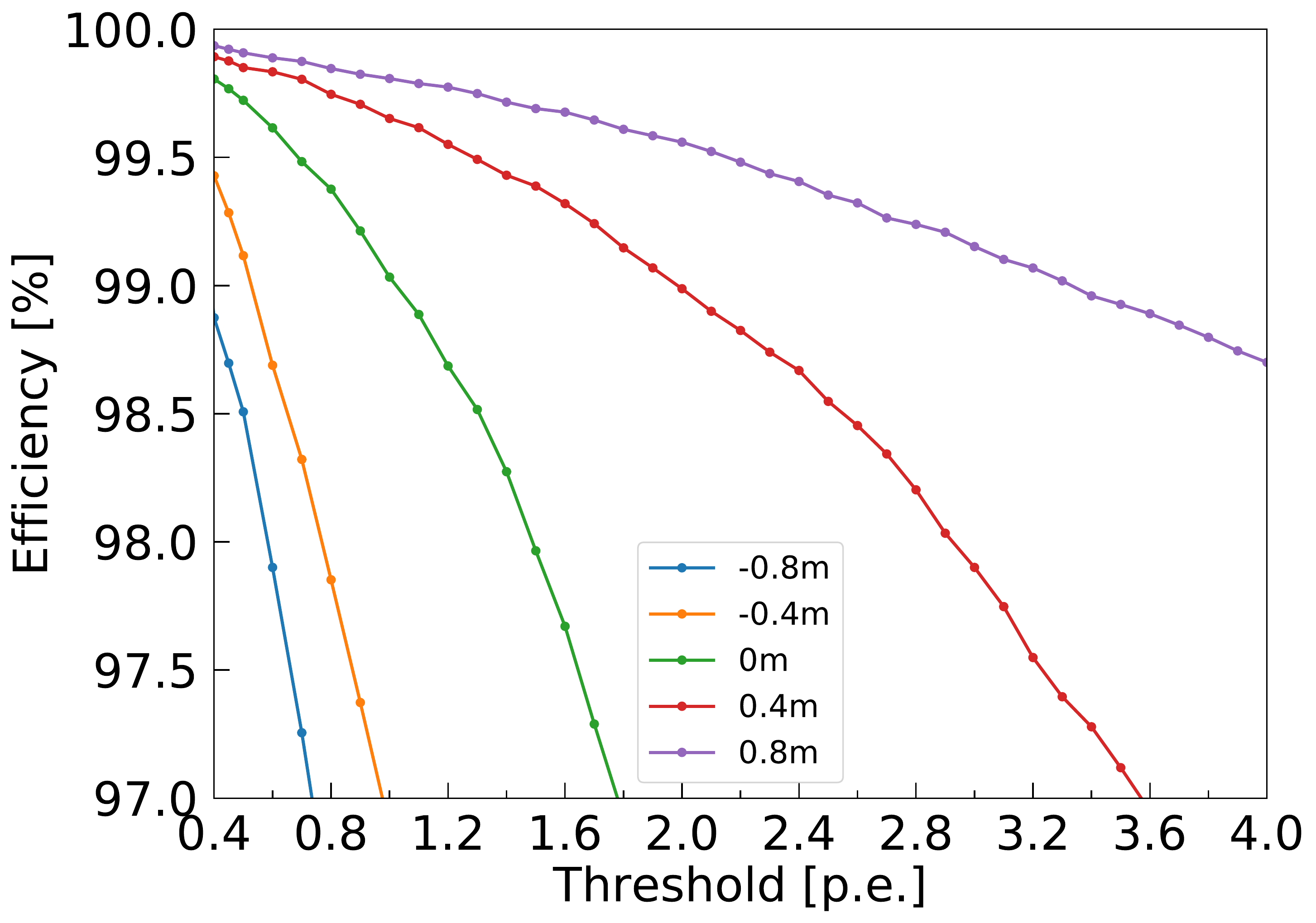}
\label{fig:spmtefficiency1}
}
\subfigure[Muon eff. vs. thrd. of double-end]{
\includegraphics[width=3.2in,height=2.4in]{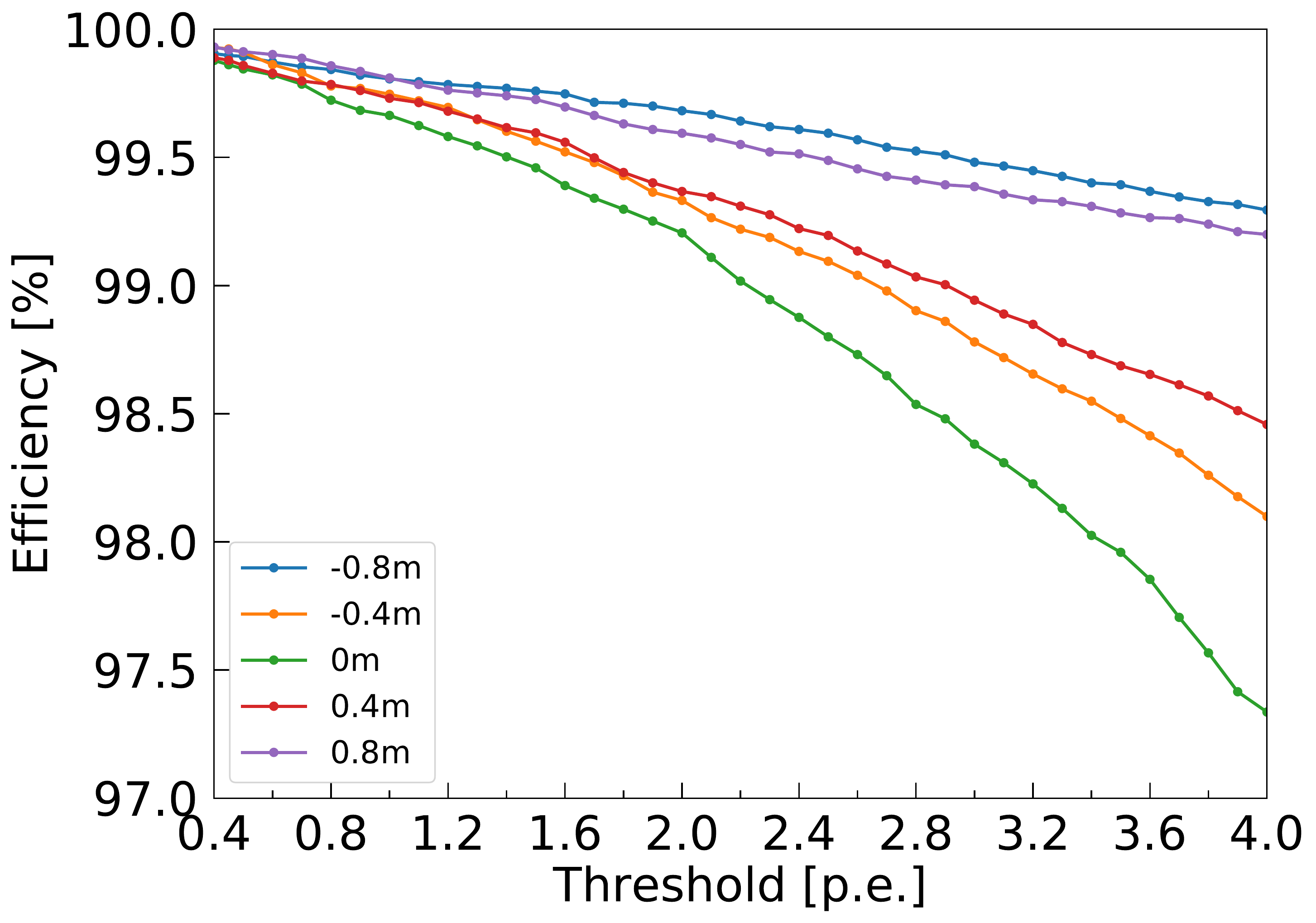}
\label{fig:spmtefficiency2}
}
\caption{Measured spectra of the measured charge spectra (top), light yield versus locations (middle), and muon efficiency versus locations (bottom) of the PS modules coupled with PMTs for single-end (left) and double-end (right) options, respectively}
\label{fig:ps:PMT} 
\end{center}
\end{figure*}

The PS strips coupled with SPMTs are measured as a reference first, where the optical window of each fiber group is further reduced to a dimension of $5\times5\,mm^2$. The two fiber groups at one end (side A or side B) of the double-end PS strip are covered by single SPMT, and each of the fiber groups of the single-end PS strip is covered by single SPMT, where no coupling gel is used between the PS and the SPMT. The charge spectra in p.e. measured by the SPMTs (sum of the two SPMTs) are plotted for both single-end and double-end options with a threshold of 0.4\,p.e., as shown in Fig.\ref{fig:spmtpecpectrum_1end} for single-end option, and Fig.\ref{fig:spmtpecpectrum_2end} for double-end option.

The light yield (LY) of each location is defined as the mean value of the measured charge spectrum. The LY of a single-end option (Fig.\,\ref{fig:spmtlightyield1}) shows a feature of monotonous decline when the muon hitting location is getting far from the readout end (-1\,m, SPMTs). In a contrast, the minimum LY of the double-end option with a symmetrical shape (Fig.\,\ref{fig:spmtlightyield2}) is at the middle of the strip as expected because of the attenuation length of the fibers and the solid angle of PS flashing point to the PMTs. The LY of the single-end option is a little bit larger than one end only of the double-end option, which is the main source of the additional light back from the far end of the fibers. 

The muon tagging efficiency of the PS strips is defined as a ratio between the event number of the coincidence of the two SPMTs over the threshold to the event number in total triggered by the mini-modules and selected by hit-time and charge as shown in Fig.\ref{fig:hittimemuonsignal} and Fig.\ref{fig:spmtchargespectrumtrigger}, and the calculated results are shown in Fig.\ref{fig:spmtefficiency1} for single-end option and Fig.\ref{fig:spmtefficiency2} for double-end option, respectively. 
With a threshold lower than 2.0\,p.e., the double-end option can reach an efficiency higher than 99\% for all the locations, but the single-end option needs to lower the threshold to around 0.5\,p.e. It is much worse for at least half of the locations with a threshold of 0.5-1\,p.e. too. 
The muon efficiency will increase similar to the double-end option if we sum the signals of the two SPMTs as a single channel of the single-end option rather than the currently used coincidence of the two SPMTs. It is workable for the option coupling with PMT, but it will introduce a higher noise from DCR without coincidence suppression for SiPM in particular.

According to the results, the double-end option is favored compared to the single-end option for better LY, the tolerance of threshold versus efficiency. It is true especially for the coupling with SiPM to suppress the higher DCR in the range of 1-3\,p.e. The measured muon rate of the double-end PS strip coupling with SPMTs can estimate with a threshold of around 10\,p.e. on the total light intensity ($\sim$4\,p.e. of each end) to reach around 95\% muon efficiency (Fig.\,\ref{fig:spmtefficiency2}) without further coincidence between strips.

\subsection{PS strip with SiPM}
\label{3:results:2:PS:sipm}

\begin{figure*}[!ht]
\begin{center}
\subfigure[Summed charge spectra]{
\includegraphics[width=3.2in,height=2.4in]{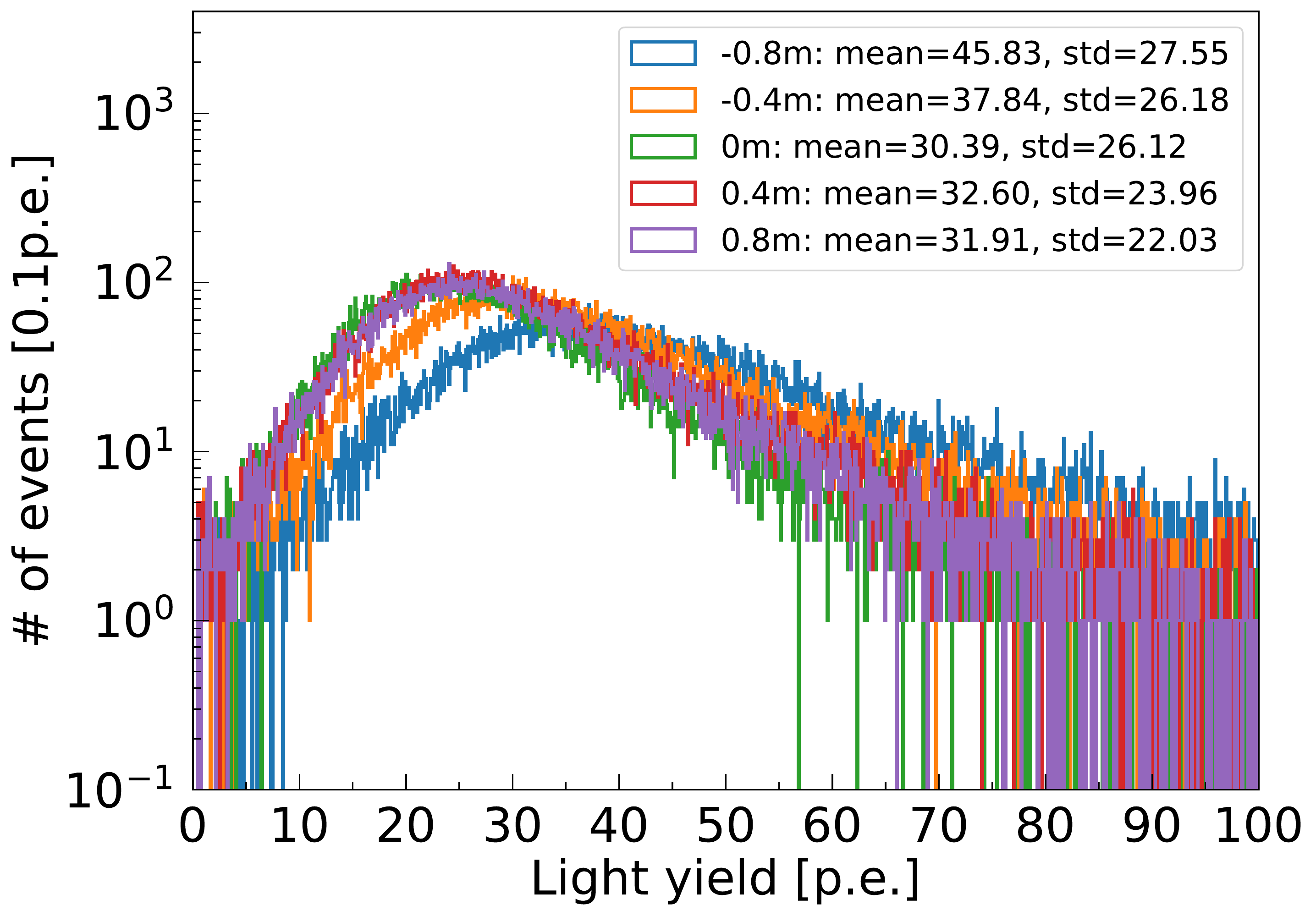}
\label{fig:sipm:charge:1}
}
\subfigure[LY vs. location]{
\includegraphics[width=3.2in,height=2.4in]{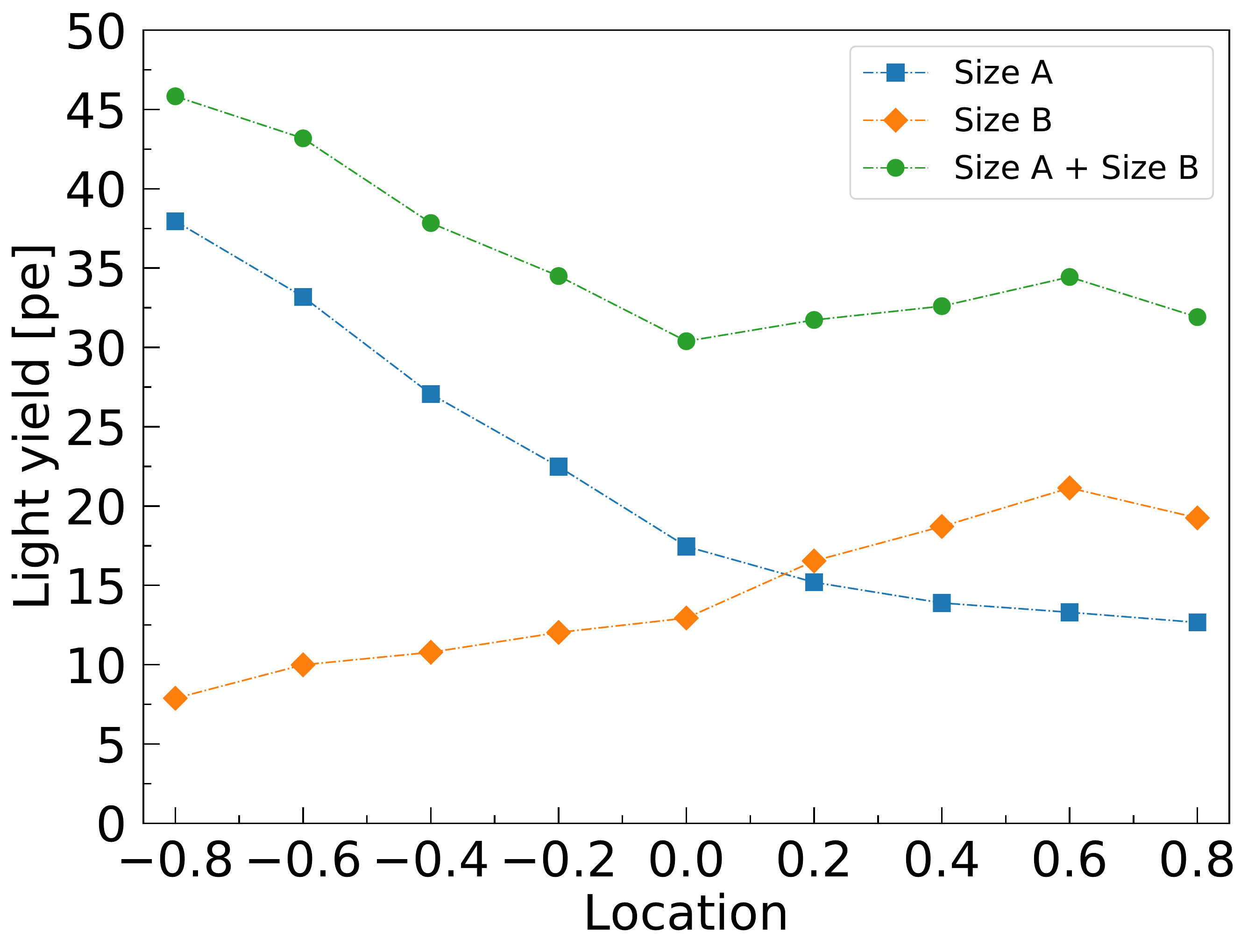}
\label{fig:sipm:pe:2}
}
\newline
\subfigure[LY vs. location by side A]{
\includegraphics[width=3.2in,height=2.4in]{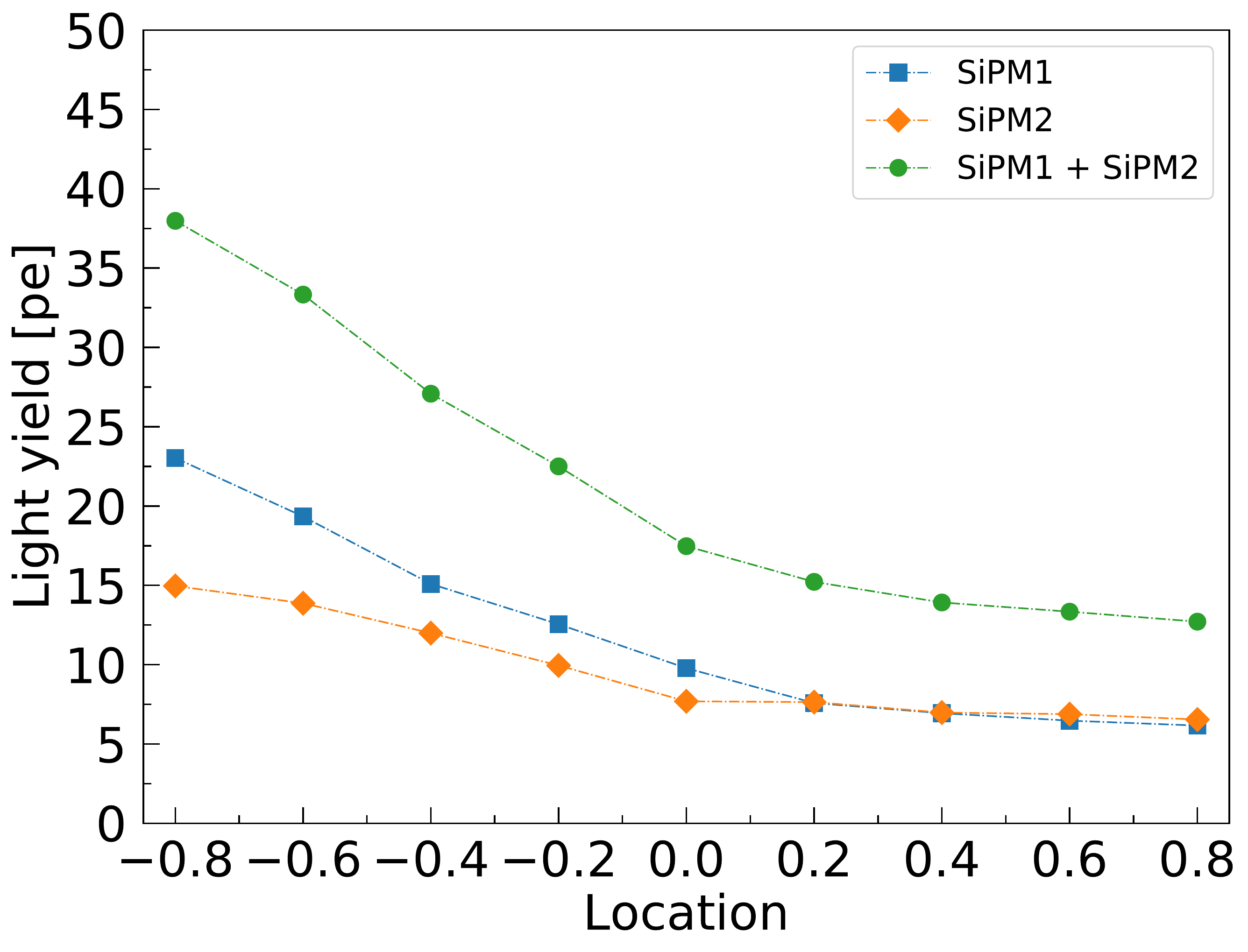}
\label{fig:sipmly:sizea}
}
\subfigure[LY vs. location by side B]{
\includegraphics[width=3.2in,height=2.4in]{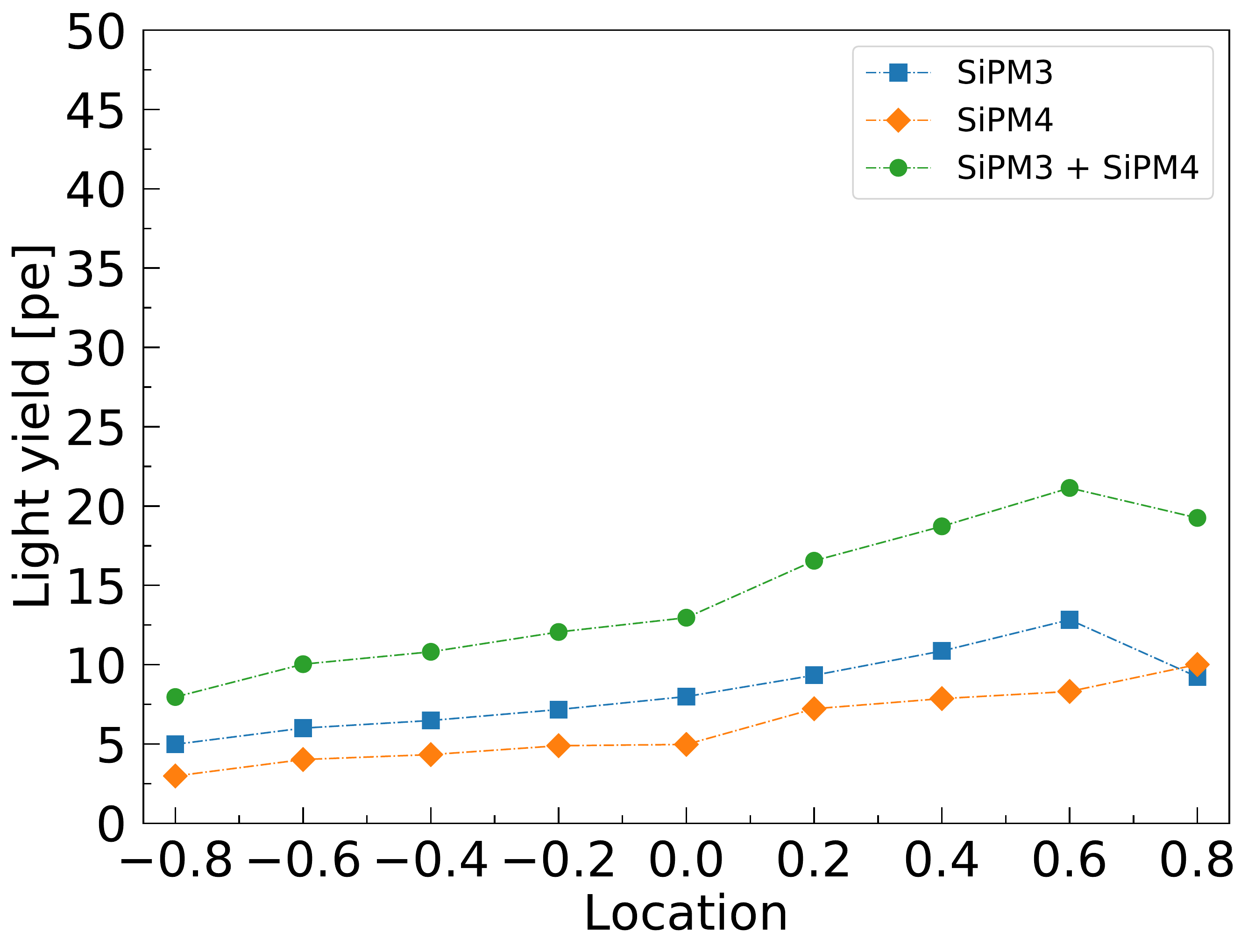}
\label{fig:sipmly:sizeb}
}
\newline
\subfigure[Muon eff. of trig. mode vs. thrd. (0\,m)]{
\includegraphics[width=3.2in,height=2.4in]{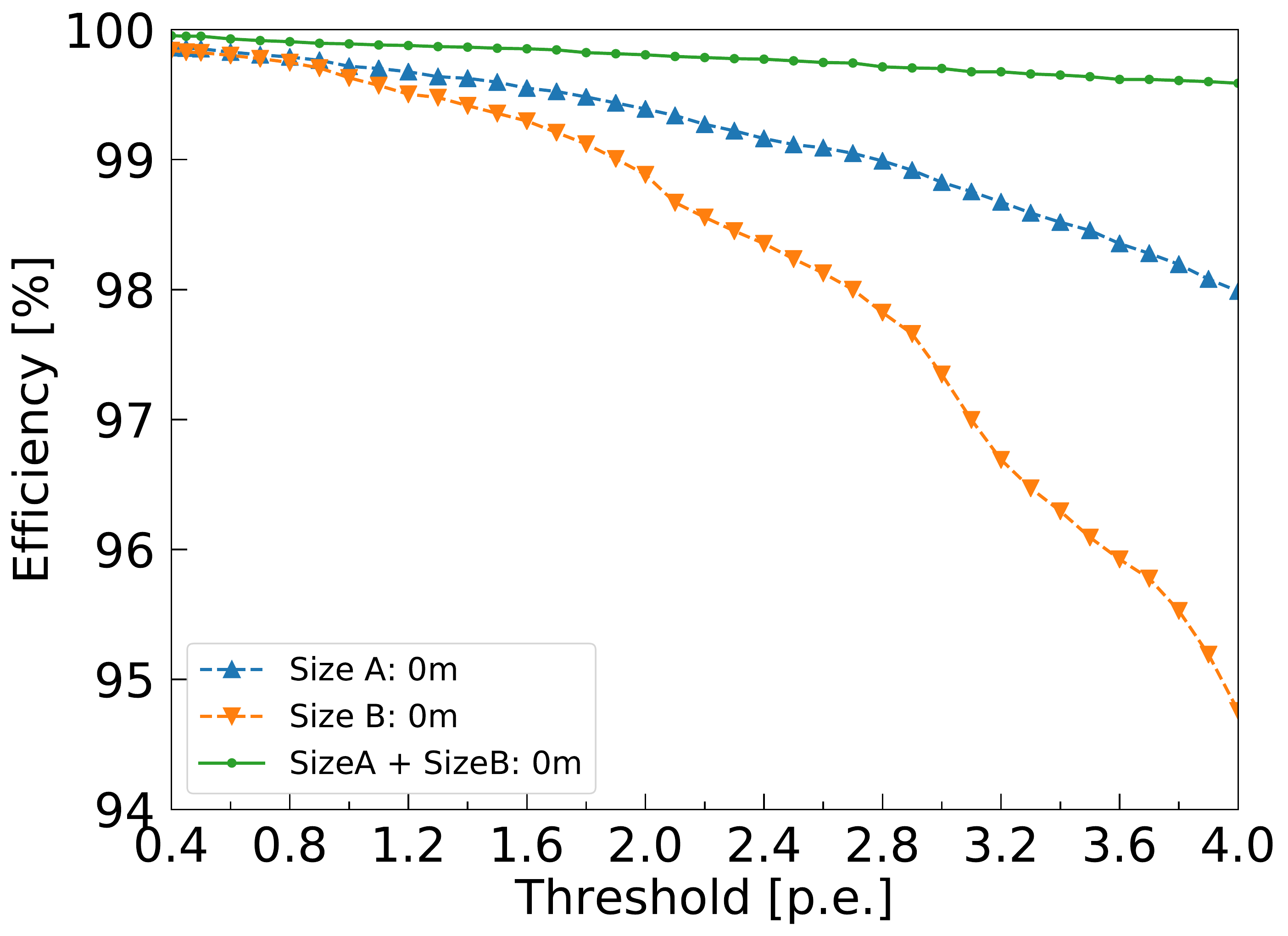}
\label{fig:muoneff:1}
}
\subfigure[Muon eff. vs. thrd. along the PS]{
\includegraphics[width=3.2in,height=2.4in]{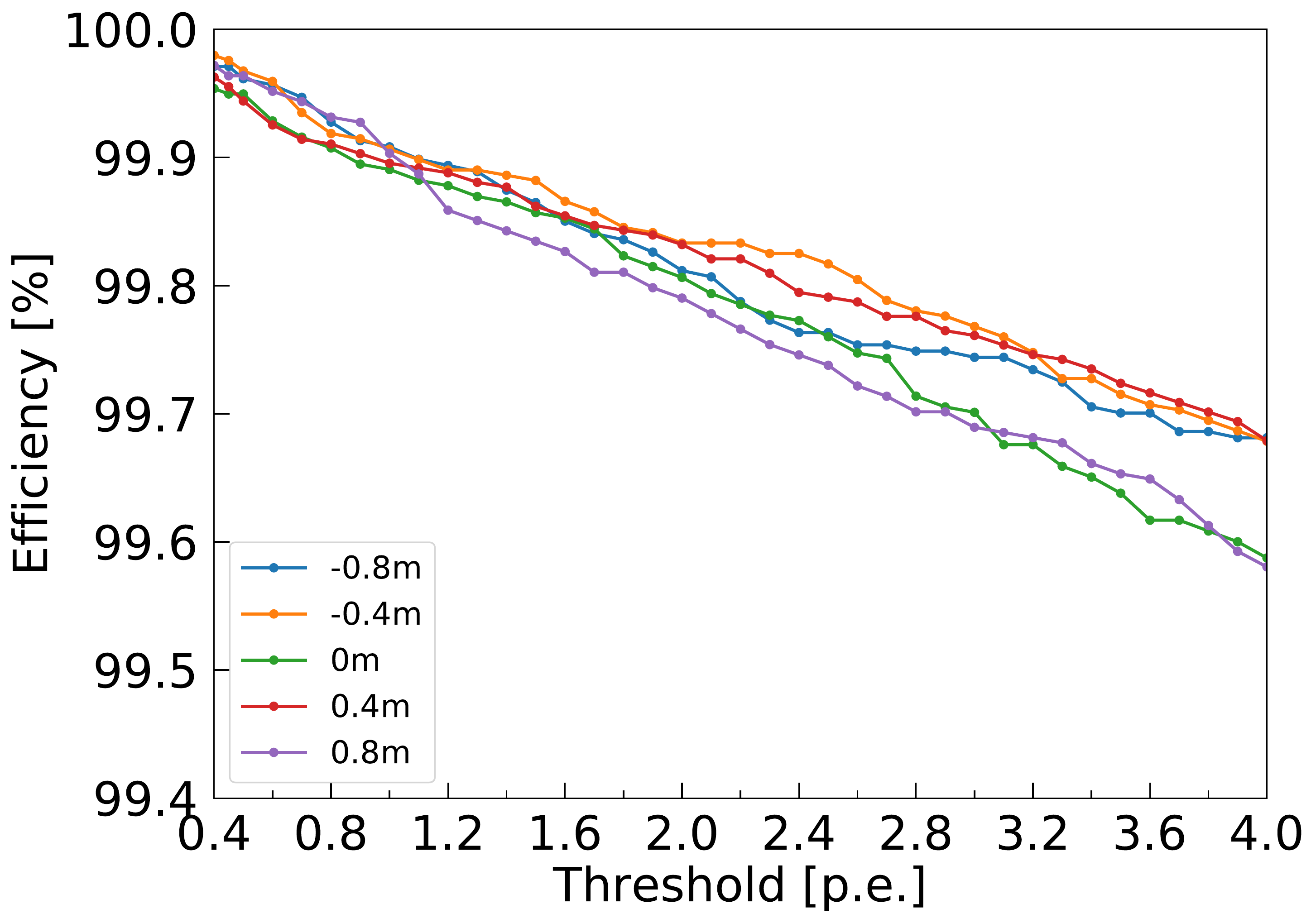}
\label{fig:muoneff:2}
}
\caption{Measured spectra of the measured charge spectra (top left), light yield in sum versus locations (top right), light yield of side A (middle left), light yield of side A (middle right), muon efficiency of location 0\,m with different trigger mode (bottom left), and muon efficiency versus different locations with triggered by both of side A and side B (bottom right), respectively.}
\label{fig:ps:SiPM} 
\end{center}
\end{figure*}

The PS strip with WLS-fiber of double-end option coupled with SiPMs is measured here, where the optical window for each fiber group is in a dimension of $5\times5\,mm^2$. Each of the optical windows is covered partially by a single SiPM cell in a dimension of $3\times3\,mm^2$ centered to the fiber, where no coupling gel is used between the PS and SiPMs. The summed spectra of the four SiPMs of the double-end option are shown in Fig.\ref{fig:sipm:charge:1}. The LY versus location is shown in Fig.\ref{fig:sipm:pe:2}, which shows a similar trend as the PS strip with SPMTs (Fig.\,\ref{fig:spmtlightyield2}), and a higher LY than that with PMTs is mainly from the QE difference and SiPM cross-talk. 
The LY inconsistency of side A and side B is further checked for each SiPM as shown in Fig.\,\ref{fig:sipmly:sizea} for side A and in Fig.\,\ref{fig:sipmly:sizeb} for side B. Both of the SiPMs in one end (SiPM1 and SiPM2, or SiPM3 and SiPM4) show a consistent trend except for the point at 0.8\,m of SiPM3. The difference in values is possible from the coupling status between PS and SiPM, the PS strip itself, and the QE of SiPMs, which still need careful validation.

The muon efficiency versus location is also derived. An example of muon efficiency with different trigger conditions at location 0\,m is shown in Fig.\ref{fig:muoneff:1}, where the efficiency of side A (also side B) is calculated only with the coincidence of SiPM1 and SiPM2 over the threshold, and the efficiency of together side A and side B is from the coincidence of side A over a threshold (sum of SiPM1 and SiPM2) and side B over a threshold (sum of SiPM3 and SiPM4). Compared with the single-end option with SPMTs as Fig.\,\ref{fig:spmtefficiency1}, the threshold can reach around 2-3\,p.e.~for higher than 99\% efficiency, but the random coincidence of SiPM DCR and the cross-talk still can not be ignored. The efficiency of together side A and side B is much better than that of the single-end as expected and discussed for the coupling with SPMT.
With the checking of all the locations, the muon efficiency of together side A and side B is shown in Fig.\,\ref{fig:muoneff:2}, and it can reach the required 99\% efficiency even with the threshold rising to around 4\,p.e., much better than the PS with PMTs, benefiting from the higher light yield even with possible SiPM coupling issues. 

Considering the threshold tolerance to the required 99\% muon efficiency, the double-end option coupled with four SiPMs is favored, which is a good choice even if there are some imperfect or unexpected issues with the PS strip, coupling, etc. A threshold of around 15\,p.e. on the sum light intensity of the double-end option coupled with four SiPMs ($\sim$7\,p.e. threshold of side A or side B) can be used to identify the muons reach an efficiency around 98-99\% without further coincidence. 

\subsection{Comparison of PS with SPMT and SiPM}
\label{3:results:3:PS:comp}

\begin{figure*}[!htb]
\begin{center}
\subfigure[Comparison of LY of PS]{
\includegraphics[width=3.2in,height=2.4in]{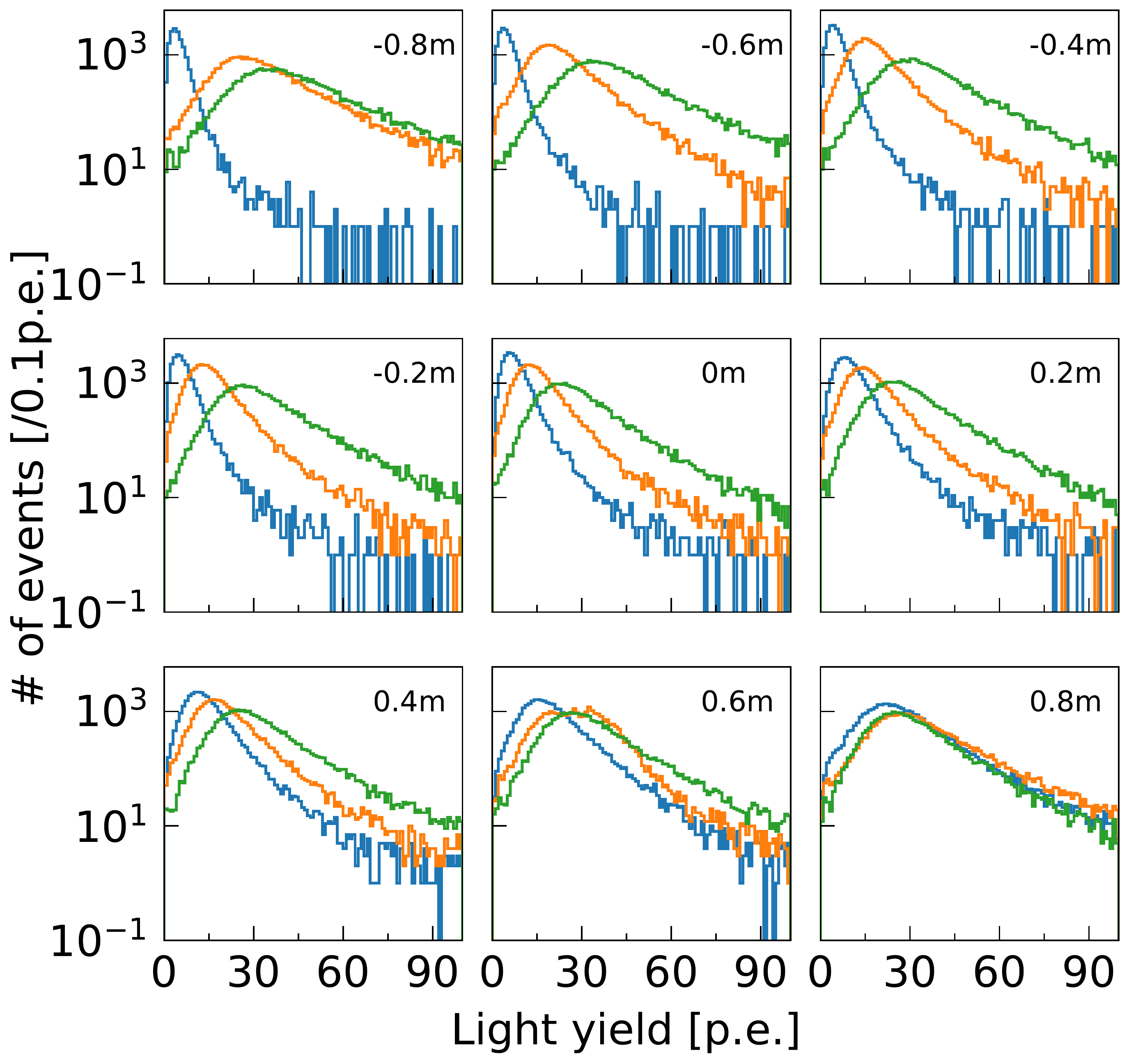}
\label{fig:comparison:pe}
}
\subfigure[Comparison of Eff. of PS]{
\includegraphics[width=3.2in,height=2.4in]{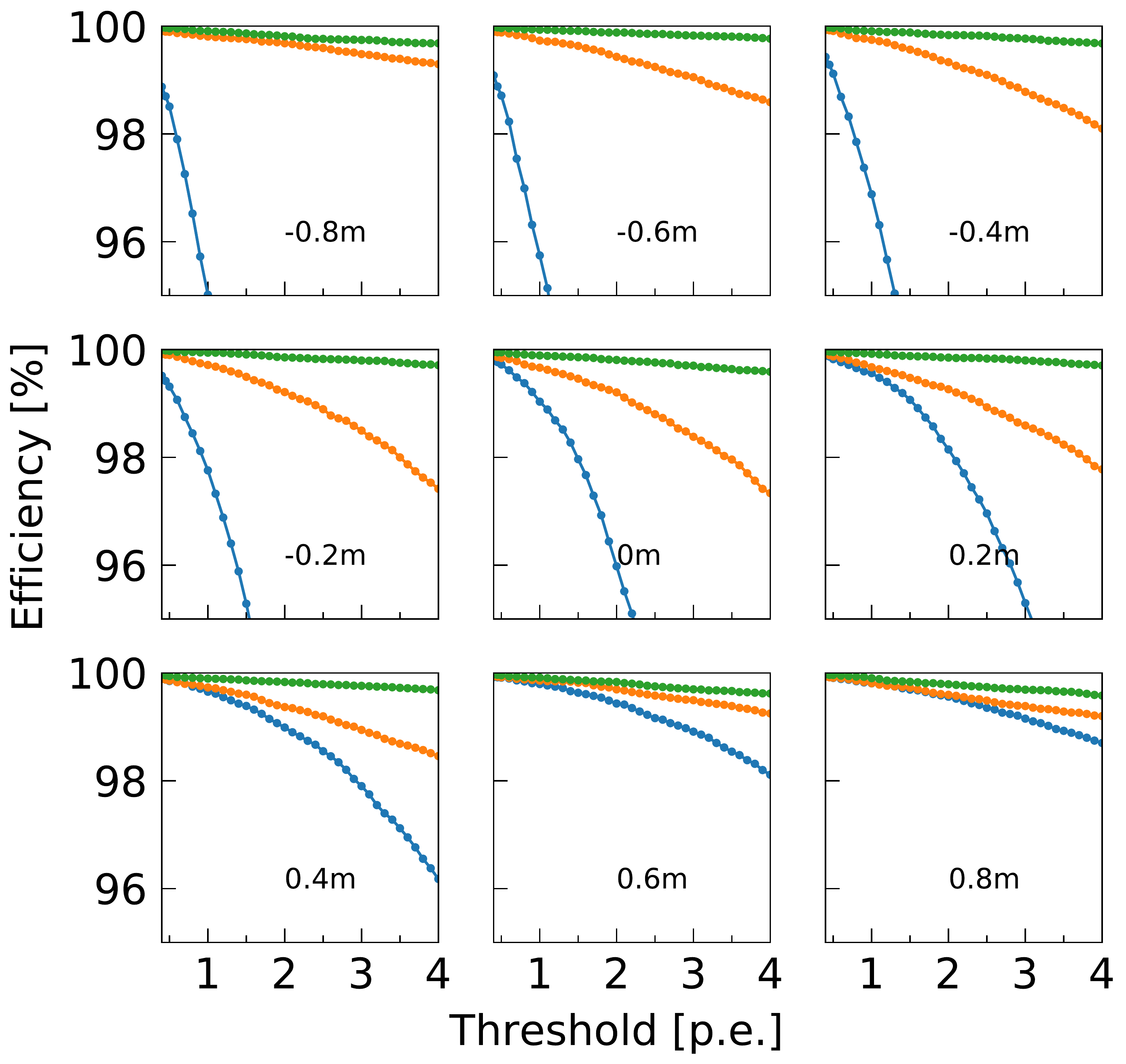}
\label{fig:comparisoneff}
}
\caption{Comparison of light yield and muon efficiency of PS strips with SiPMs and PMTs. green: double-end (SiPM), orange: double-end (SPMT), blue: single-end (SPMT).}
\label{fig:comparisonsipmandpmt} 
\end{center}
\end{figure*}

A direct comparison among the PS strips of single-end, and double-end options is shown in Fig.\ref{fig:comparisonsipmandpmt}, where the double-end option shows a higher light yield (Fig.\ref{fig:comparison:pe}) and better muon efficiency (Fig.\ref{fig:comparisoneff}) under the same threshold for both coupling with SPMT and SiPM. 
The option coupled with SiPM has the highest LY compared to the other two options, even at the location around 0.8\,m (near the end of side B), where the SiPMs or the coupling meet some issues. And the efficiency of the option coupled with SiPM under the same threshold is also highest than the other two options.

Concerning the readout options between the PMT and SiPM of double-end, the SiPM is suggested for higher light yield, efficiency, and reachable threshold. The SiPM option can also minimize the occupied dimension for a similar active volume as designed. The higher random coincidence rate suffered from the DCR of SiPM can be suppressed by the coincidence of side A and side B or by increasing the threshold (55\,kHz/mm$^2$, 100\,ns coincidence window, 12\% cross-talk, (SiPM1 or SiPM2) and (SiPM3 or SiPM4)): $\sim$98\,kHz at threshold 0.5\,p.e., $\sim$1.5\,kHz at threshold 1.5\,p.e., $\sim$24\,Hz at threshold 2.5\,p.e., and $\sim$0.37\,Hz at threshold 3.5\,p.e. The temperature effect of the SiPM option can also be minimized by the high LY and reachable threshold.

\begin{figure*}[!htb]
\begin{center}
\subfigure[Comparison of LY of PS]{
\includegraphics[width=3.2in,height=2.4in]{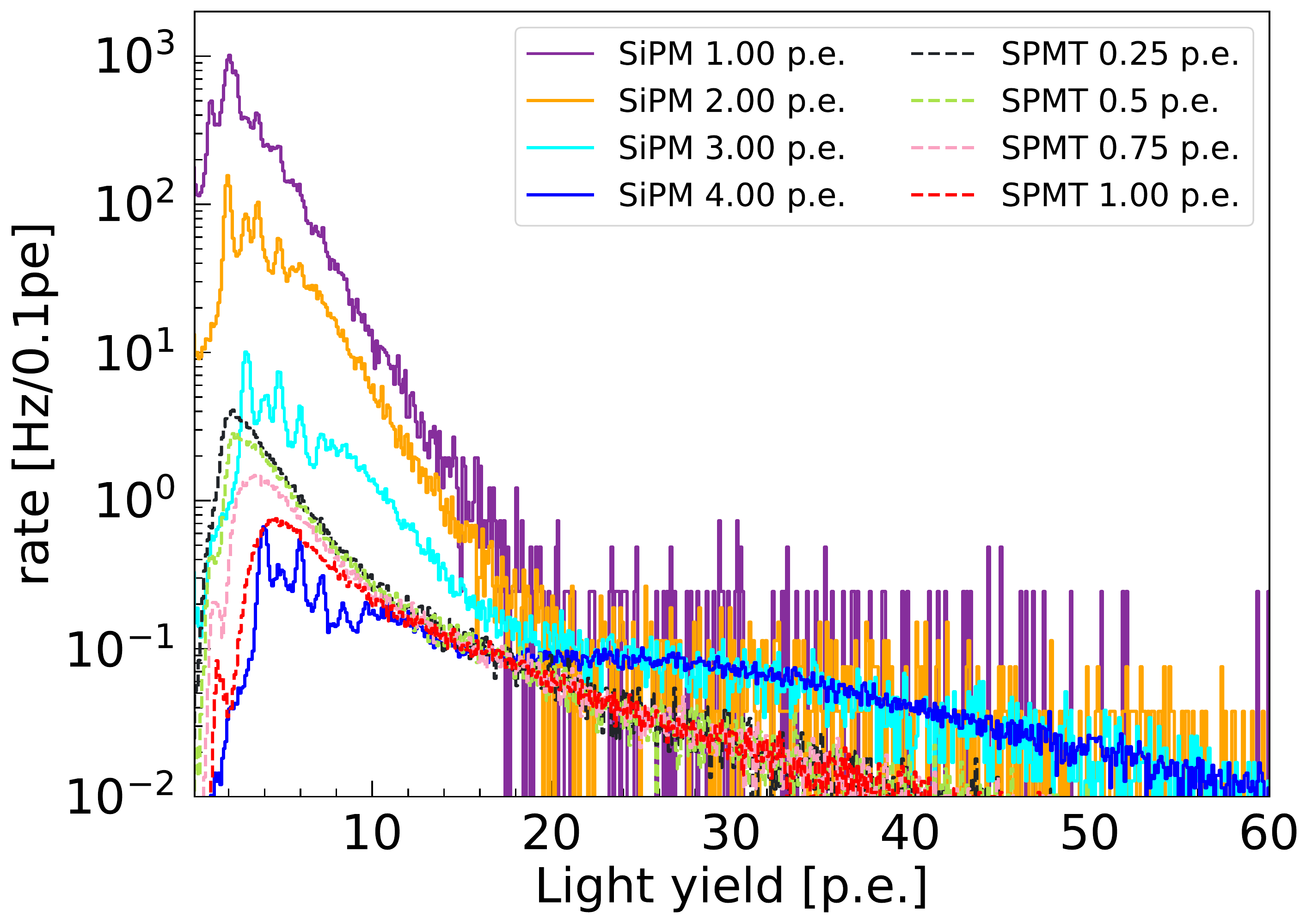}
\label{fig:PS:full:spectra:charge}
}
\subfigure[Comparison of Eff. of PS]{
\includegraphics[width=3.2in,height=2.4in]{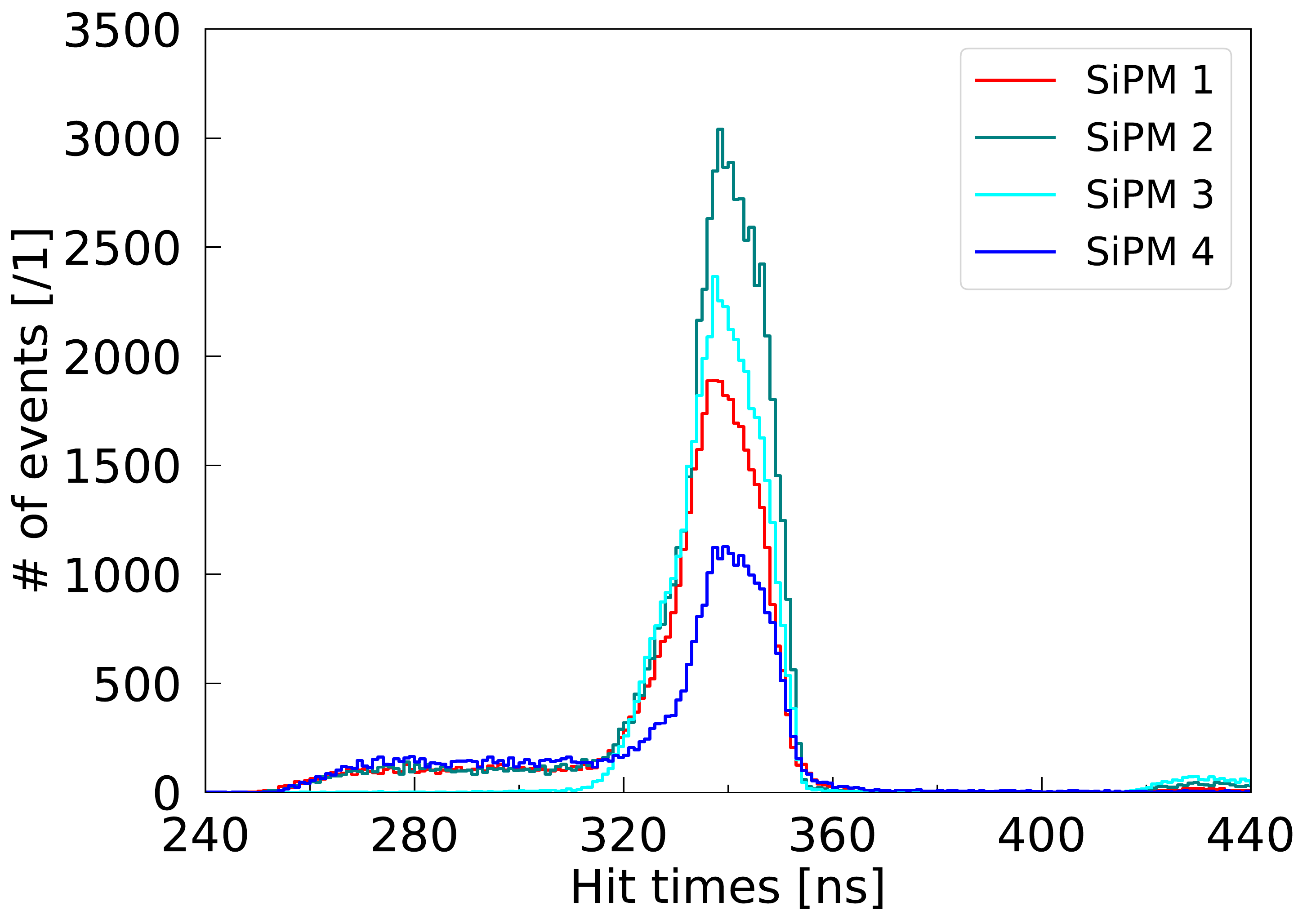}
\label{fig:PS:full:spectra:hittime}
}
\caption{Comparison of light yield and muon efficiency of PS strips with SiPMs and PMTs. green: double-end (SiPM), orange: double-end (SPMT), blue: single-end (SPMT)}
\label{fig:PS:full:spectra}
\end{center}
\end{figure*}

A comparison of the directly measured charge spectra by the PS strip of double-end option coupled with SPMT and SiPM is shown in Fig.\ref{fig:PS:full:spectra:charge}, the Y axis is normalized to the muon event of the coupling of SiPM or SPMT, and the measured rate between the coupling of SiPM and SPMT. The threshold is 0.25\,p.e., 0.5\,p.e., 0.75\,p.e. and 1.0\,p.e. for each SPMT with a coincidence of side A and side B, and 1.0\,p.e., 2.0\,p.e., 3.0\,p.e. and 4.0\,p.e. for each SiPM with a coincidence of SiPM1 and SiPM2 (side A) or SiPM3 and SiPM4 (side B), which is not the coincidence of side A (sum of SiPM1 and SiPM2) and side B (sum of SiPM3 and SiPM4) limited by the hardware. The expected muon efficiency of all configurations is higher than 98-99\%. 

The measured rate of the strip coupled with SPMTs is $\sim$160\,Hz, $\sim$120\,Hz, $\sim$80\,Hz and $\sim$50\,Hz with each threshold setting as listed, respectively, which mainly contributed by the natural radioactivity and cosmic-muon (around 20\,Hz under the muon rate of 100\,Hz$/m^2$ at sea level), and the random coincidence of the SPMT DCR is less than 0.1\,Hz. It is around 20\,Hz higher than the 10\,p.e. threshold of muon to the sum of side A and side B, and the relative contribution of the radioactivity is around 140\,Hz for 0.25\,p.e. threshold. The measured rate of the strip coupled with SiPMs is $\sim$30,000\,Hz, $\sim$4,500\,Hz, $\sim$320\,Hz and $\sim$50\,Hz of each threshold setting as shown in Tab.\,\ref{table:rate}, respectively. It is around 20\,Hz higher than 15\,p.e. threshold as muon selection to the charge sum of side A and side B. 

\begin{table}[!htb]
\centering
\caption{Coincidence rate of the double-end strip coupled with SPMT and SiPM with different threshold.}
\label{table:rate}
\begin{tabular}{c|c|c|c|c} 
 \hline
 Threshold (p.e.) & & & & \\
 SPMT & 0.25 & 0.5 & 0.75 & 1.0 \\
 SiPM & 1.0 & 2.0 & 3.0 & 4.0 \\
 \hline\hline
 SPMT Rate/Hz  & $\sim$160 & $\sim$120 & $\sim$80 & $\sim$50 \\ 
 \hline
 SiPM Rate/Hz & $\sim$30,000 & $\sim$4,500 & $\sim$320 & $\sim$50 \\
 \hline
 SiPM Random & & & \\
 Coincidence Rate/Hz & $>$10,730 & $>$2,280 & $>$180 & $>$15 \\ [1ex] 
 \hline
\end{tabular}
\end{table}

The measured rate of the PS strip coupled with SiPM is much higher than that coupled with SPMT, the lower threshold in particular, which is contributed partially by the radioactivity with the higher LY and mainly by the random coincidence of DCR. There are lots of events located in the region where it is lower than the required sum light intensity of all the SiPMs over the threshold, including lower than 2\,p.e.~of 1\,p.e.~threshold, lower than 4\,p.e. of 2\,p.e.~threshold, lower than 6\,p.e.~of 3\,p.e.~threshold, and lower than 8\,p.e.~of 4\,p.e.~threshold. The reason is from the required hit time region in [320,360]\,ns as SiPM analysis window considering for the difference of muon hit time. But it is 100\,ns of the coincidence window of the two SiPMs of side A or side B for data acquisition trigger, where the random coincidence of SiPM DCR will contribute a lot as an example shown in Fig.\,\ref{fig:PS:full:spectra:hittime}. In this case, only one of the hits inside the analysis window from SiPM DCR, while the others out of the analysis window will be ignored, will be taken into account for the summed light intensity calculation and contribute to the region which is higher than the threshold of single SiPM but lower than the sum of their coincidence. In another hand, we can calculate the random coincidence rate from SiPM DCR with this region of the summed light intensity spectra roughly, which is correlated and proportional to the real random coincidence. The calculated random coincidence rate is around $\sim$
10,730\,Hz, $\sim$2,280\,Hz, $\sim$180\,Hz and $\sim$15\,Hz of each SiPM threshold setting (Tab.\,\ref{table:rate}), which is basically consistent with the previous calculation.

\section{Summary}
\label{4:summary}

In this article, a compact design of the plastic scintillator with WLS-fiber is proposed, fabricated, and measured. The options are compared between single-end and double-end, and coupling with PMT and SiPM. The results demonstrate that the double-end option is a good choice for JUNO-TAO considering the light yield and muon efficiency. The option coupled with SiPM achieves a better performance along 2\,m plastic scintillator, and better tolerance to noise and threshold, even though the coupling of one side meets some issues (The coupling between PS and SiPM can be improved further.). The proposed option with SiPM is workable for a compact requirement, an appreciable light yield (minimum $\sim$30\,p.e./muon) and robust. It offers a quantitative candidate of scintillator/WLS-fiber configuration for future muon tagging detectors. Furthermore, the results can also be referred to for other general designs of scintillator-based detectors with WLS-fiber read-out.

\begin{acknowledgements}
This work was supported partially by the National Natural Science Foundation of China (Grant No. 11875282 and 12022505), the Strategic Priority Research Program of the Chinese Academy of Sciences (Grant No. XDA10011200), and the Youth Innovation Promotion Association of CAS.
\end{acknowledgements}

\bibliographystyle{unsrt}
\bibliography{allcites}   

\end{document}